\documentclass[11pt]{article}

\usepackage{yfonts,epsfig}
\usepackage{subfigure}
\usepackage{supertabular}
\usepackage{amsmath}
\usepackage{amssymb}
\usepackage{cite}
\usepackage[sans]{dsfont}
\usepackage{graphicx}
\usepackage[dvips]{color}
\usepackage{soul}
\usepackage{comment}
\usepackage{lscape}
\usepackage{caption}
\usepackage{appendix}
\usepackage{CJK}

\newcommand{\clfnt}{\setcounter{footnote}{0}}
\newcommand{\cleqn}{\setcounter{equation}{0}}
\newcommand{\newc}{\newcommand}
\newc{\om}{\omega}
\newc{\im}{{\bf i}}
\newc{\sig}{\sigma}
\newc{\epi}{\epsilon}
\newc{\beq}{\begin{equation}}
\newc{\eeq}{\end{equation}}
\newc{\beqn}{\begin{eqnarray}}
\newc{\eeqn}{\end{eqnarray}}
\newc{\bsym}{\boldsymbol}
\newc{\ol}{\overline}
\newc{\unl}{\underline}
\newc{\bmat}{\begin{pmatrix}}
\newc{\emat}{\end{pmatrix}}
\def\bea{\begin{eqnarray}}
\def\eea{\end{eqnarray}}
\def\bean{\begin{eqnarray*}}
\def\eean{\end{eqnarray*}}
\def\m#1{\mathcal#1}
\def\B#1{\textbf#1}

\newcommand{\qed}{\nobreak \ifvmode \relax \else
      \ifdim\lastskip<1.5em \hskip-\lastskip
      \hskip1.5em plus0em minus0.5em \fi \nobreak
      \vrule height0.75em width0.5em depth0.25em\fi}

\newcommand{\be}{\begin{equation}}
\newcommand{\ee}{\end{equation}}
\newcommand{\barr}{\begin{array}}
\newcommand{\earr}{\end{array}}

\newcommand{\bed}{\begin{displaymath}}
\newcommand{\eed}{\end{displaymath}}

\renewcommand{\t}{\tilde}

\DeclareMathOperator{\Tr}{Tr}

 \def\bvec#1{\raise1.5ex\hbox{$\rightarrow$}\mkern16.5mu #1}

 \def\m#1{\mathcal#1}
\begin{document}


\begin{titlepage}
\begin{flushright}\today
\end{flushright}
\vspace{1cm}

\begin{center}
{\bf \Large On Mixing Supersymmetry and Family Symmetry Breakings}
\end{center}
\vskip 1cm

\begin{center}{
{\bf M. Jay P\'erez${}^{\,a}_{}$, Pierre Ramond${}^{\, b}_{}$,  and Jue Zhang\begin{CJK}{GBK}{}\end{CJK}${}^{\,c}_{}$
}
\vskip .9cm
{\em  Institute for Fundamental Theory,\\
Department of Physics, University of Florida\\
Gainesville FL 32611, USA}
\vskip .2cm

}\end{center}

\vskip .9cm
\begin{abstract}
\noindent  
We present a toy model in which the Higgs sector fields transform as non-Abelian representations of a family symmetry group, and consider the possibility that the extra family partners of the Higgs particles act as messengers for both supersymmetry and family symmetry breakings. Although such mediation schemes generically produce family dependent soft supersymmetry breaking terms at the messenger scale, we demonstrate the existence of a focusing mechanism which may erase such hierarchies through renormalization group running. 
\end{abstract}

\vskip 1.4cm


\vspace{1cm}
\vfill \vskip 5mm \hrule width 5.cm \vskip 2mm {\small
\noindent $^a~$mjperez@ufl.edu\\
\noindent $^b~$ramond@phys.ufl.edu\\
\noindent $^c~$juezhang@ufl.edu
}
\end{titlepage}




\section{Introduction}
\cleqn 

Although the Standard Model (SM) has passed all experimental scrutiny, it is not without its puzzles.  One is the unnaturally small mass of the Higgs boson. Unprotected by any symmetry, it should be of the order of the cutoff of the theory. A promising candidate for such a protective symmetry is supersymmetry. Another puzzle is the number of chiral families, which is suggestive of a family symmetry. Finally, while the quark and lepton gauge quantum numbers suggest their (Grand) unification, their peculiar mass patterns and mixings are very different. 
 
If nature indeed chooses a supersymmetric extension of the SM, one question that continues to bewilder is the ``$\mu$ problem'' of the Minimal Supersymmetric Standard Model (MSSM) (see review \cite{Martin:1997ns}).  Phenomenology constrains $\mu$ to be of the order of weak scale, while, being a supersymmetric invariant, one would naturally expect it to be of the order of the cutoff, the Grand Unified Theory (GUT) or Plank scale. Is there a symmetry to suppress or forbid the $\mu$ term?  

Family symmetry would be a natural candidate, although most models consider the electroweak doublet Higgs fields to be family singlets. If family symmetry protects the $\mu$ term, the Higgs would necessarily transform under some representation of the family group and have extra family partners. A natural question is what role, if any, these extra particles may play. 

In  minimal gauge mediation of supersymmetry breaking \cite{Giudice:1998bp}, there are heavy fields with Higgs-like gauge quantum numbers that act as messengers of supersymmetry breaking. They are family singlets to avoid excessive Flavor Changing Neutral Current (FCNC) effects.

The purpose of this paper is to see if these Higgs family partners can act as the messengers which bring both supersymmetry and family symmetry breakings to the MSSM. Since they have family quantum numbers, the soft supersymmetry breaking terms generated will be family dependent. We examine whether such an approach is capable of avoiding potentially lethal FCNC effects.  

To study this question, we consider a toy model where the MSSM includes $\m S_3$ as a family symmetry.  The breakings of family symmetry and supersymmetry occur through the vacuum expectation values of spurion fields. These spurions couple to Higgs family multiplets, which then transmit the breakings to the MSSM. 

Physical consequences depend on the pattern of spurion vacuum values. For all vacuum configurations, there exists regions of parameter space  where electroweak symmetry breaking occurs at tree-level, but this leads to unsustainable sum rules characteristic of tree-level supersymmetry breaking. 

In some regions of parameter space, tree-level symmetry breaking is avoided; the structure of the soft terms is driven by radiative corrections generated through Yukawa and gauge interactions. Unlike gauge contributions, the Yukawa interactions generate soft terms at one-loop. In some configurations, they generically give rise to negative soft masses squared for the sfermion fields, resulting in unphysical electroweak symmetry breaking.

Fortunately, there exists other vacuum configurations for which the lowest order term in the one-loop Yukawa contributions vanishes. In this case, all soft terms  arise at two-loops, and are family dependent.  We find a mechanism by which, through renormalization group (RG) running, family hierarchies in the ultraviolet can be ameliorated in the infrared. This arises because the family symmetry correlates boundary values with the parameters of the running. This opens up the intriguing possibility of models in which flavored mediation is not as dangerous as once thought.

This $\m S_3$ model has shortcomings;  it includes only two families, and a $\mu$-like term. It does not contain color triplet Higgs messengers (which would be present in a GUT generalization) so that the gluinos are massless. 

There is scarce literature on this subject  \cite{Chacko:2001km,Evans:2010kd,Evans:2011bea,Evans:2012hg,Shadmi:2011hs,Abdullah:2012tq,
Kang:2012ra,Craig:2012xp,Craig:2012yd,Albaid:2012qk}, and although previous authors have considered Higgs-messenger mixing, none have considered models where the messengers are family partners of the Higgs fields. We hope this model may serve as an exemplar for further model building aimed at studying these questions.  

Our paper is organized as follows.  In section 2, we present the model. We explore the case where the dominant contributions to the soft terms occur at one-loop in Section 3.  Section 4 and 5 contain a description of vacuum alignments for which the two-loop contributions dominate, as well as the RG running of the soft parameters. We outline there the conditions necessary for family dependence in the boundary values to be mitigated. Finally, we summarize the properties of this model and outline plans for future work in section 6. Technical details are relegated to the appendices throughout.

\section{$\m S_3$ Model}
\cleqn

In order to illustrate our ideas, we consider a supersymmetric model invariant under $\m S_3$, the smallest non-Abelian discrete group. $\m S_3$ is the permutation group on three objects. Its presentation is given by

\begin{equation}
\langle a , b | a^3 = b^2 = e, bab^{-1}=a^{-1} \rangle .
\end{equation}
It contains three irreducible representations: two singlets $\mathbf{1}$ and $\mathbf{1}^\prime$, and  a doublet $\mathbf{2}$. We choose a basis such that the primed singlet and the two dimensional irreducible representations of our group elements are given by

\begin{eqnarray}
\B 1^\prime &:& ~~ a = 1, ~~ b = -1, \nonumber \\
\B 2 &:& ~~ a = \begin{pmatrix} \omega & 0 \\ 0 & \omega^2 \end{pmatrix} , \qquad b =  \begin{pmatrix}  0 & 1 \\ 1 & 0 \end{pmatrix},
\end{eqnarray}
where $\omega = e^{2 \pi i / 3}$. 
The relevant Kronecker products are

\begin{eqnarray}
\mathbf{1}^\prime \otimes \mathbf{2} &=& \mathbf{2}, \nonumber \\
\mathbf{2} \otimes \mathbf{2} &=& \mathbf{1}_a^\prime \oplus (\mathbf{1} \oplus \mathbf{2})_s, 
\end{eqnarray}
and the Clebsch-Gordan (CG) coefficients are given by

\begin{eqnarray}
x_{1'} \otimes \begin{pmatrix} y_1 \\ y_2 \end{pmatrix} &=& \begin{pmatrix} -x y_1 \\ x y_2 \end{pmatrix}, \\
\begin{pmatrix} x_1 \\ x_2 \end{pmatrix} \otimes \begin{pmatrix} y_1 \\ y_2 \end{pmatrix} &=& (x_1 y_2 + x_2 y_1)_1 \oplus (x_1 y_2 - x_2 y_1)_{1^\prime} \oplus \begin{pmatrix} x_2 y_2 \\ x_1 y_1 \end{pmatrix}_2.
\end{eqnarray}
The  visible superpotential consists of two parts: a Yukawa sector that describes two chiral families, with fields transforming as $\m S_3$ doublets,    

\be
W_Y = y_uQ\bar{u}\m H_{u}+y_dQ\bar{d}\m H_{d}+y_eL\bar{e}\m H_{d}
+ \t y_u Q \bar{u} \phi_u + \t y_d Q \bar{d} \phi_d + \t y_e L \bar{e} \phi_d,
\ee  
which gives rise to the following family structure:

\bea \label{eqn:yukawa}
W_Y &=& y_u(Q_1\bar{u}_1\m H_{u1}+Q_2\bar{u}_2\m H_{u2})+y_d(Q_1\bar{d}_1\m H_{d1}+Q_2\bar{d}_2\m H_{d2})\nonumber \\ &+& y_e(L_1\bar{e}_1\m H_{d1}+L_2\bar{e}_2\m H_{d2}) +  \t y_u (Q_1 \bar{u}_2 - Q_2 \bar{u}_1)\phi_u \nonumber \\
&+& \t y_d (Q_1 \bar{d}_2 - Q_2 \bar{d}_1)\phi_d + \t y_e (L_1 \bar{e}_2 - L_2 \bar{e}_1)\phi_d,
\eea
and a Higgs sector, 

\begin{eqnarray}
W_{\m H} &=& m \m H_u \m H_d +M \phi_u  \phi_d + \lambda \Delta \m H_u \m H_d +\t \lambda_u \Delta \phi_u \m H_d + \t \lambda_d \Delta \phi_d \m H_u.
\end{eqnarray}
The fields $\phi_u$ and $\phi_d$ are electroweak doublets which transform as the $\B 1'$ representation of $\m S_3$; $\Delta$ is a SM singlet and $\m S_3$ doublet, and all other fields are $\m S_3$ doublets. A summary of the field content of the model is given in Table \ref{tb:fields}.  Although $m$ looks like a $\mu$ term, its role is quite different in this simple model.

The $\phi_{u,d}$ fields are necessary to reduce the global symmetries of the superpotential to $\m S_3$, along with total lepton and baryon numbers and a $U(1)$ $R$-symmetry. \footnote{If not included, the symmetry group of the superpotential is much larger, including two lepton and two baryon numbers, one for each family.}   

Assuming $M$ very large decouples $\phi_u$ and $\phi_d$ from the low energy theory, and generates non-renormalizable terms in the  superpotential,

\begin{eqnarray}
\label{eq:nonren}
W_{nonren} &=& \frac{1}{M} \Bigg( \t \lambda_u \t \lambda_d (\m H_u \Delta)_{\B 1^\prime} (\m H_d \Delta)_{\B 1^\prime} \nonumber \\
& & + \t y_u \t \lambda_d (Q \bar{u})_{\B 1^\prime} (\m H_u \Delta)_{\B 1^\prime} +  \t y_d \t \lambda_u (Q \bar{d})_{\B 1^\prime} (\m H_d \Delta)_{\B 1^\prime}  + \t y_e \t \lambda_u (L \bar{e})_{\B 1^\prime} (\m H_d \Delta)_{\B 1^\prime} \nonumber \\
& & +\t y_u \t y_d (Q \bar{u})_{\B 1^\prime} (Q \bar{d})_{\B 1^\prime} + \t y_u \t y_e (Q \bar{u})_{\B 1^\prime} (L \bar{e})_{\B 1^\prime} \Bigg)
+ \m O \left( \frac{1}{M^2} \right).
\end{eqnarray}
The renormalizable part of the superpotential reduces to

\begin{equation}
W_{ren} = m \m H_u \m H_d +  \lambda \Delta \m H_u \m H_d + y_uQ\bar{u}\m H_{u}+y_dQ\bar{d}\m H_{d}+y_eL\bar{e}\m H_{d}.
\end{equation}

We assume that $\Delta$ is the only field in the visible superpotential that couples to a hidden sector. Hidden sector dynamics are unknown (to us), but we assume that it leaves $\Delta$ with a vacuum configuration that breaks {\em both} family symmetry and supersymmetry, that is 

\be
\langle \lambda\Delta \rangle= v\psi +\theta^2  F \psi',
\ee
where $\psi$ and $\psi'$ are constant family-space doublets; $v$ and $F$ describe the strengths of family and supersymmetry breakings, respectively.  There are many different vacua, each giving rise to distinct phenomenology. We focus on two possibilities which extremize the invariants, 

$$\psi=\begin{pmatrix}
\psi_1 \\ \psi_2
\end{pmatrix},~~~\psi'=\begin{pmatrix}
\psi_1' \\ \psi_2'
\end{pmatrix}~~\sim ~~\begin{pmatrix}
0 \\ 1
\end{pmatrix},    \quad \begin{pmatrix}
1 \\  1
\end{pmatrix}.$$

\begin{table}[!h]
\begin{center}
\begin{tabular}{|c||c|c|c|c|c|c|c|c|c|c|}
\hline 
\hline
$ $ & $Q$ & $\bar{u}$ & $\bar{d}$ & $L$ & $\bar{e}$ & $\m H_u$ & $\m H_d$ & $\phi_u$ & $\phi_d$ & $\Delta$ \\
\hline
$\m S_3$ & $\B 2$ & $\B 2$ & $\B 2$ & $\B 2$ & $\B 2$ & $\B 2$ & $\B 2$ & $\B 1^\prime$ & $\B 1^\prime$ & $\B 2$    \\
\hline
$U(1)_R$ & $1/2$ & $1/2$ & $1/2$ & $1/2$ & $1/2$ & $1$ & $1$ & $1$ & $1$ & $0$ \\
\hline 
\hline
\end{tabular}
\caption{Field content of the model.}
\label{tb:fields}
\end{center}
\end{table}

In the following analyses, we will assume that the scale at which the family group is broken is larger than that of supersymmetry breaking, that is

$$F\ll v^2\ll M^2,$$
while keeping $m< v$, but otherwise undetermined. 
After replacing $\Delta$ with its $vev$ and integrating out the $\phi_{u,d}$ fields, the quadratic part of the superpotential becomes,

\be
\m H_u^t \B M \m H_d +\theta^2 \m H_u^t \B F \m H_d,
\ee
where 

\be
\label{eq:matrices}
\B M=\begin{pmatrix} v\psi_1 & m\\ m & v\psi_2
\end{pmatrix},\qquad \B F=F\begin{pmatrix} \psi_1' &0\\ 0 & \psi_2'
\end{pmatrix},
\ee
are $(2 \times 2)$ matrices in family space determined from the CG coefficients and vacuum directions. We take the parameters $m$, $v$, and $F$ to be real and positive. The diagonal entries are approximate, as integrating out the $\phi_{u,d}$ fields will give contributions from 

\begin{equation}
\label{eq:lift}
\frac{1}{M}\m H_{u 1} \m H_{d 1}(v\psi_2+F\theta^2\psi_2')^2,\qquad  \frac{1}{M}\m H_{u 2} \m H_{d 2}(v\psi_1+F\theta^2\psi_1')^2,
\end{equation}
but such contributions can be neglected if $M$ is large enough. 

We now proceed to a detailed analysis of the model for several vacuum configurations. 

\section{Aligned Vacuum}
\clfnt 
\cleqn
We choose the vacuum structure where family and supersymmetry breaking are aligned,

$$\psi=\psi^\prime=\begin{pmatrix} 0 \\  1\end{pmatrix},$$ 
so that 

\be
\bf M=  \begin{pmatrix}
0 & m \\ m & v
\end{pmatrix},\qquad \bf F=  \begin{pmatrix}
0 & 0 \\ 0 & F
\end{pmatrix}.
\ee
One nice property of the supersymmetric mass matrix $\B M$ is that it is of the ``see-saw'' form: if $m \ll v$, we have a large hierarchy between the two pairs of Higssinos. The scalar potential is\footnote{The $SU(2)$ component structure is shown for clarity, using $\epsilon^{1 2} = - \epsilon_{1 2} = 1$. A lower index represents the doublet representation, while an upper index is its complex conjugate.}

\begin{eqnarray}
V_\m H &=& 
\begin{pmatrix} \m {H_{u}}^{\dag~\alpha} & \m {H_{d~\beta}}^{t}\epsilon^{\beta \alpha}\end{pmatrix}
\begin{pmatrix} \B M^\dag \B M  & \B F \\
\B F & \B M^\dag \B M \end{pmatrix}
\begin{pmatrix}  \m {H_{u~\alpha}}\\
\epsilon_{\alpha \gamma}\m {H_d}^{*~\gamma}\end{pmatrix} \nonumber \\
~ &+& \frac{1}{8}g^{\prime 2}(\m H_u^{\dagger} \m H_u - \m H_d^{\dagger} \m H_d)^2 +\frac{1}{8}g^2(\m H_u^{\dagger} \tau^a \m H_u + \m H_d^{\dagger} \tau^a \m H_d)^2 ,
\end{eqnarray}
where $g^\prime$ and $g$ are the $U(1)$ and $SU(2)$ gauge couplings, and $\tau^a/2$ are the $SU(2)$ generators. 

The combinations

\be
\epsilon^{\alpha \beta}\Phi_{+~\beta} = \frac{1}{\sqrt{2}}(\epsilon^{\alpha \gamma}\m {H_{u~\gamma}}+\m {H_d}^{*~\alpha}),\qquad 
\Phi_-^{*~\alpha} = \frac{1}{\sqrt{2}}(-\epsilon^{\alpha \beta}\m {H_{u~\beta}}+\m {H_d^{*~\alpha}}),
\ee
yield mass squared matrices in block diagonal forms

\begin{eqnarray}
V_{quad.}=\begin{pmatrix} \Phi_+^{\dag~\gamma} \epsilon_{\gamma \beta} & \Phi_{-~\beta}^{t}\end{pmatrix}
\begin{pmatrix} \B M^\dag \B M+\B F & 0\\
0 & \B M^\dag \B M-\B F
\end{pmatrix}
\begin{pmatrix}\epsilon^{\beta \delta} \Phi_{+~\delta}\\
\Phi_-^{*~\beta} \end{pmatrix} .
\end{eqnarray}
The diagonalisation of each block yields the mass eigenstates $H_+$ and $H_-$, by performing rotations $\Phi_+=\m R(\theta_+) H_+$ and $\Phi_-=\m R(\theta_-) H_-$ in family space, with

\begin{eqnarray}
\label{eq:transformation}
\begin{pmatrix} \m {H_u}\\ \m {H_d}^*\end{pmatrix} &=&
\frac{1}{\sqrt{2}}\begin{pmatrix} \m R(\theta_+) & -\m R(\theta_-)\\ \m R(\theta_+) &
\m R(\theta_-)\end{pmatrix}
\begin{pmatrix} H_+\\
H_-^*\end{pmatrix}, \qquad
\m R(\theta_{\pm}) = \begin{pmatrix} \cos{\theta_\pm} & -\sin{\theta_\pm}\\
\sin{\theta_\pm} & \cos{\theta_\pm}\end{pmatrix},
\end{eqnarray}
resulting in the diagonalized scalar potential,

\begin{eqnarray} \label{eq:diagpot}
V_{H} &=& m_{+ 1}^2 |H_{+ 1}|^2 + m_{+ 2}^2 |H_{+ 2}|^2 + m_{- 1}^2 |H_{- 1}|^2 + m_{- 2}^2 |H_{- 2}|^2 \nonumber \\
~ &+& \frac{1}{32}g^{\prime 2}(H_+^{\dagger} \m R^{t} (\bar{\theta}) H_-^*+h.c.)^2+\frac{1}{32}g^2(H_+^{\dagger} \m R^{t} (\bar{\theta}) \tau^a H_-^*+h.c.)^2,
\end{eqnarray}
where the eigenvalues and mixing angles are given by 

\begin{eqnarray}
\label{eq:eigen}
m_{\pm 1}^2 &=& m^2 + \frac{1}{2} (v^2\pm F) - \frac{1}{2} \sqrt{(v^2\pm F)^2+4m^2v^2} , \nonumber
\\
m_{\pm 2}^2 &=& m^2 + \frac{1}{2} (v^2\pm F) + \frac{1}{2} \sqrt{(v^2 \pm F)^2+4m^2v^2} , \nonumber
\\
\tan (2\theta_{\pm}) &=& \frac{2mv}{v^2 \pm F}  , \qquad \bar{\theta} = \theta_+- \theta_- , \qquad \theta_{\pm} \in [0,\frac{\pi}{2}].
\end{eqnarray}
The relations  

\begin{eqnarray}
\label{eq:trdet}
\det(\B M^\dagger\B M\pm \B F) &=& m^2(m^2\pm F), \\
\Tr(\B M^\dagger\B M\pm \B F) &=& 2m^2+v^2 \pm F > 0,
\end{eqnarray}
show that the signs of the squared masses depend on the relative size of $F$ and $m^2$: if $m^2 < F$ one of the four mass squared is negative, while for $m^2 >F$ all mass squared are positive.

\subsection{$m^2 < F$}

If $m^2<F \ll v^2$, $H_{- 1}$ has negative mass squared,  indicating tree-level spontaneous symmetry breaking. The potential of Eq.(\ref{eq:diagpot}) is unbounded from below along  $\langle H_{- 2} \rangle = 0$ because of a flat direction, but is lifted by the interactions Eq.(\ref{eq:nonren}),\footnote{Further lifting from the term $\Delta \m H_u \m H_d$ is present when we treat $\Delta$ as a dynamical field.} yielding tree-level electroweak symmetry breaking. The altered potential is,

\begin{eqnarray}
V_H^\prime &=& V_H + \eta |\m H_{u1}\m H_{d1}|^2 \\ \nonumber
&=& V_H + \frac{\eta}{4} \Big ( \big |[ \m R(\theta_+)H_+]_1 \big |^2 - \big | [\m R(\theta_-)H_-]_1 \big |^2  \Big )^2 
-\frac{\eta}{4} \Big ( [\m R(\theta_+)H_+]_1 [\m R(\theta_-)H_-]_1 +h.c \Big)^2
\end{eqnarray}
where $\eta=\t \lambda_u \t \lambda_d v/M$ is a dimensionless parameter that accounts for the interactions Eq.(\ref{eq:nonren}) generated by the $\phi_{u,d}$ fields. Since $m_{+1,2}^2$ are both positive, we set at the minimum,

\bea
\langle H_{+1,2} \rangle = 0, \qquad 
\langle H_{-1} \rangle = \begin{pmatrix}
v_1 \\ 0
\emat, \qquad \langle H_{-2} \rangle = \begin{pmatrix}
v_2 \\ 0
\end{pmatrix},
\eea
the latter determined from  $\partial V_{H}^\prime / \partial H_{-1}^- =0$. The $vev$'s  $v_1$ and $v_2$ for $m^2 < F \ll v^2$ are given by,

\be
 v_1 \approx \frac{m}{v} \sqrt{\frac{2(F-m^2)}{\eta}} ,\qquad v_2 \approx \frac{m^3 F}{v^5} v_1,
\ee
which indicates that electroweak symmetry breaking is mostly triggered by $H_{-1}$. 

In this vacuum, mass terms are generated for the various fields by their couplings to the $H_-$ fields:

\begin{itemize}

\item All sfermions receive tree-level masses except for the sneutrinos.

\item In the Higgs sector we are left with four massive CP-even scalars, three massive CP-odd scalars, and three massive complex charged scalars. 

\item All fermions receive tree-level masses except for the neutrinos and gluinos. The gluinos, which would ordinarily obtain mass terms radiatively, remain massless in the absence of color triplet messengers.

\end{itemize}
We have included a detailed calculation of the spectrum in Appendix A.  

As supersymmetry breaking is transmitted only by renormalizable, tree-level couplings, the supertrace theorems constrain the resulting spectrum of masses, leading to well known phenomenological problems; the average mass of the sfermion fields is around that of their fermion superpartners. These constraints may be avoided if our model is extended to include supergravity, as this will add terms to the supertrace gravitino mass. As long as the gravitino mass is not too light ($\sim$ TeV), an acceptable spectrum may still be possible. 

\subsection{$m^2 > F$}

When $m^2 > F$, all masses squared are positive and there is no tree-level spontaneous symmetry breaking. The masses of the Higgs fields are hierarchical, 

\begin{eqnarray}
\label{eq:Higgs_masses}
m_{\pm 1}^2 &= & \frac{m^2}{v^2}\left(  m^2 \pm F + \ldots \right) \nonumber \\
m_{\pm 2}^2 &= &  v^2\left(1+2\left(\frac{m}{v}\right)^2 \pm \frac{F}{v^2} + \ldots \right),
\end{eqnarray}
where we have neglected terms of $\m O(1/v^4)$. From the rotation angles,

\be
\theta_{\pm} = \frac{m}{v} \mp \frac{m F}{v^3} + \m O(1/v^3),
\ee
the Yukawa couplings become,

\begin{eqnarray}
\m L_Y &\supset & y_uQ_1\bar{u}_1\left[H_u-\frac{m}{v}M_u \right ]+y_uQ_2\bar{u}_2\left[\frac{m}{v} H_u+M_u \right ] \nonumber \\
& & +y_dQ_1\bar{d}_1\left[H_d-\frac{m}{v} M_d \right ]+y_dQ_2\bar{d}_2\left[\frac{m}{v}H_d+M_d \right ] \nonumber \\
& & +y_eL_1\bar{e}_1\left[H_d-\frac{m}{v}M_d \right ]+y_eL_2\bar{e}_2\left[\frac{m}{v} H_d+M_d  \right ]+ \m O\left( 1/v^3 \right),
\end{eqnarray}
written in terms of the combinations (not mass eigenstates) of light ($H_{u,d}$) and heavy  ($M_{u,d}$) fields,

\begin{eqnarray}
H_u = \frac{1}{\sqrt{2}}\left( H_{+1}-H_{-1}^*\right), \qquad H_d = \frac{1}{\sqrt{2}}\left( H_{+1}^*+H_{-1}\right), \nonumber \\
M_u = \frac{1}{\sqrt{2}}\left( H_{+2}-H_{-2}^*\right), \qquad M_d = \frac{1}{\sqrt{2}}\left( H_{+2}^*+H_{-2}\right). \nonumber
\end{eqnarray}
A Yukawa hierarchy emerges. The second families effective Yukawa coupling is suppressed by $(m/v) \ll 1$, relative to the first family. However in this simplified model, the hierarchical structure is the same for the up quarks, down quarks and charged leptons. 

The fermions remain massless since electroweak symmetry breaking has not yet taken place and we must call on radiative corrections to achieve it. In addition, as the Higgs fields are now the mediators of supersymmetry breaking, some soft supersymmetry breaking terms are expected to be generated radiatively.


\subsubsection{Gaugino masses}
\clfnt

As in ordinary gauge mediation, gaugino masses are generated at the one-loop level. The masses of bino and winos are given by \cite{Marques:2009yu} (gluinos are massless in this simple model):

\begin{eqnarray} \label{eqn:gauginomass}
M_{1} &=&\frac{g^{\prime 2}}{16\pi^2} \Lambda_G, \qquad M_{2} = \frac{g^2}{16\pi^2} \Lambda_G,  \\ \nonumber
\Lambda_G &=& \sum_{i,j=1,2; \pm} \pm \left |\m R(\theta_\pm -\theta)^\dagger_{ij}\right |^2 \frac{ m_{0j} m_{\pm i}^2 \ln (m_{\pm i}^2/m_{0j}^2)}{m_{\pm i}^2-m_{0j}^2},
\end{eqnarray}
where $m_{0i}$ are the eigenvalues of the supersymmetric mass matrix $\B M$, and $\theta$ is the angle of the rotation matrix which diagonalizes $\B M$. Their values can be obtained by setting $F=0$ in Eq.(\ref{eq:eigen}). Note that the above  mass formulae are only true when all masses squared are positive, which requires the constraint $m^2>F$.

We are interested in the limit $F \ll v^2$ of Eq.(\ref{eqn:gauginomass}). Defining as dimensionless expansion parameters $x \equiv m/v$ and $y \equiv F/v^2$, we have two extreme cases. The first is when $x \ll 1$ which we think of as the limit where $m \ll v$. The other is when $x \lesssim 1$, which indicates that $m$ is close to $v$.  

\begin{enumerate}

\item
When $x \ll 1$, depending on the relative magnitudes of $y$ and $x^2$, we have the following two expansions of $\Lambda_G/({F/v})$,

\begin{eqnarray} \label{eq:gauginoexpansion}
0<y\ll x^2 \ll 1 : &~&  \frac{\Lambda_G}{F/v} = 2+\frac{1}{6}\left(\frac{y}{x^2}\right)^2-4x^2+\m O\left(\left(\frac{y}{x^2}\right)^4, x^4, x^2y^2 \right), \nonumber \\
0<y\lesssim x^2 \ll 1 : &~&  \frac{\Lambda_G}{F/v} = 1+2\ln 2+(3\ln 2-1)\left(1-\frac{y}{x^2}\right)+\left(1-\frac{y}{x^2}\right)\ln\left(1-\frac{y}{x^2}\right)\nonumber \\ 
&~&~~~~~~~~~~ -4x^2+\m O\left(x^4, \left(1-\frac{y}{x^2}\right)^2\right),
\end{eqnarray}
where we have used $(1-y/x^2)$ as one of the expansion parameters.

Compare this formula to that of the ordinary gauge mediated picture with $N$ decoupled set of messengers,

\begin{equation}
M_a = \sum_{i=1}^n \frac{g_a^2}{16 \pi^2} \frac{\m F_i}{\m M_i} =  N_{eff} \frac{g_a^2}{16 \pi^2} \frac{\m F}{\m M},
\end{equation}
where $\m M_i$ and $\m F_i$ are masses and supersymmetry breaking $F$ $vev$'s of messengers ($\m F_i \ll \m M_i^2$) and $N_{eff}$ is called ``effective number of messengers''. 

Eq.(\ref{eq:gauginoexpansion}) indicates the effective number of messengers to be twice as big as naively expected for only one source of supersymmetry breaking $F$. This is because the light Higgs fields also contribute to the gaugino masses through the (11) entry of the following matrix,

\begin{equation} \label{eqn:fdiag}
\m R(\theta) \B F \m R^t(\theta) = \begin{pmatrix}
F \sin^2{\theta} & - F \sin{\theta} \cos{\theta} \\ 
 - F \sin{\theta} \cos{\theta} & F \cos^2{\theta}
\end{pmatrix}.
\end{equation}

Although the supersymmetry breaking parts for the first family are much smaller than those of the heavy Higgs fields (recall $\sin{\theta} \approx x$), what is really important is their contribution in a given loop diagram, which goes as $\m F_i / \m M_i$.  Although the effective $F$ term for the light fields is suppressed by $x^2$, their masses are also suppressed by the same ratio, $m_{light} \approx x^2 v$, and $\m F_i / \m M_i \approx F / v$. Such contributions are thus comparable with those from the heavy Higgs fields, resulting in the above expansion results.

\item
If $0< y \ll x \lesssim 1$, we find 

\begin{eqnarray}
\frac{\Lambda_G}{F/v} &=& \frac{2\sqrt{5}}{5}+\frac{8\sqrt{5}}{25}(1-x)+\left( \frac{53\sqrt{5}}{750}+\frac{2}{125}\ln \left(\frac{7}{2}-\frac{3\sqrt{5}}{2}\right)\right) y^2 \nonumber \\
& & +\m O((1-x)^2,(1-x)y^2,y^4),
\end{eqnarray}
and the effective number of messengers is approximately 1. The mixing between the Higgs fields is now quite large, and $F$ is roughly split between the two Higgs families. This is compensated by a splitting of the masses, which when 

\begin{equation}
\B M \approx \begin{pmatrix} 0 & v \\ v & v \end{pmatrix},
\end{equation}
gives eigenvalues $m_{0~{1,2}} = (\sqrt{5} \mp  1 ) v / 2$. Using Eq.(\ref{eqn:fdiag}), we can approximate the effective number of messengers as

\begin{equation}
\sum_{i=1}^2 \frac{\m F_i}{\m M_i} \approx \left( \frac{F\sin^2{\theta}}{m_{01}} + \frac{F \cos^2{\theta}}{m_{02}} \right) \approx  \frac{F}{v} \left( \frac{2\sqrt{5}}{5} \right) ,
\end{equation}
which gives an effective number of messengers of $0.89$. 
\end{enumerate}

For intermediate values of our parameters we turn to a numerical analysis. The top two graphs of  Fig.\ref{fg:gaugino} give the  dependence on $y$ for $x=0.15$ and $x=1$, showing that $\Lambda_G/(F/v)$ depends weakly on $y$. In the bottom graph, with fixed $y$ one sees that with increasing $x$, $\Lambda_G/(F/v)$ decreases from 2 to 1, agreeing with the above analysis. 

\begin{figure}[h!]
 \centering
\scalebox{0.35}{\includegraphics*{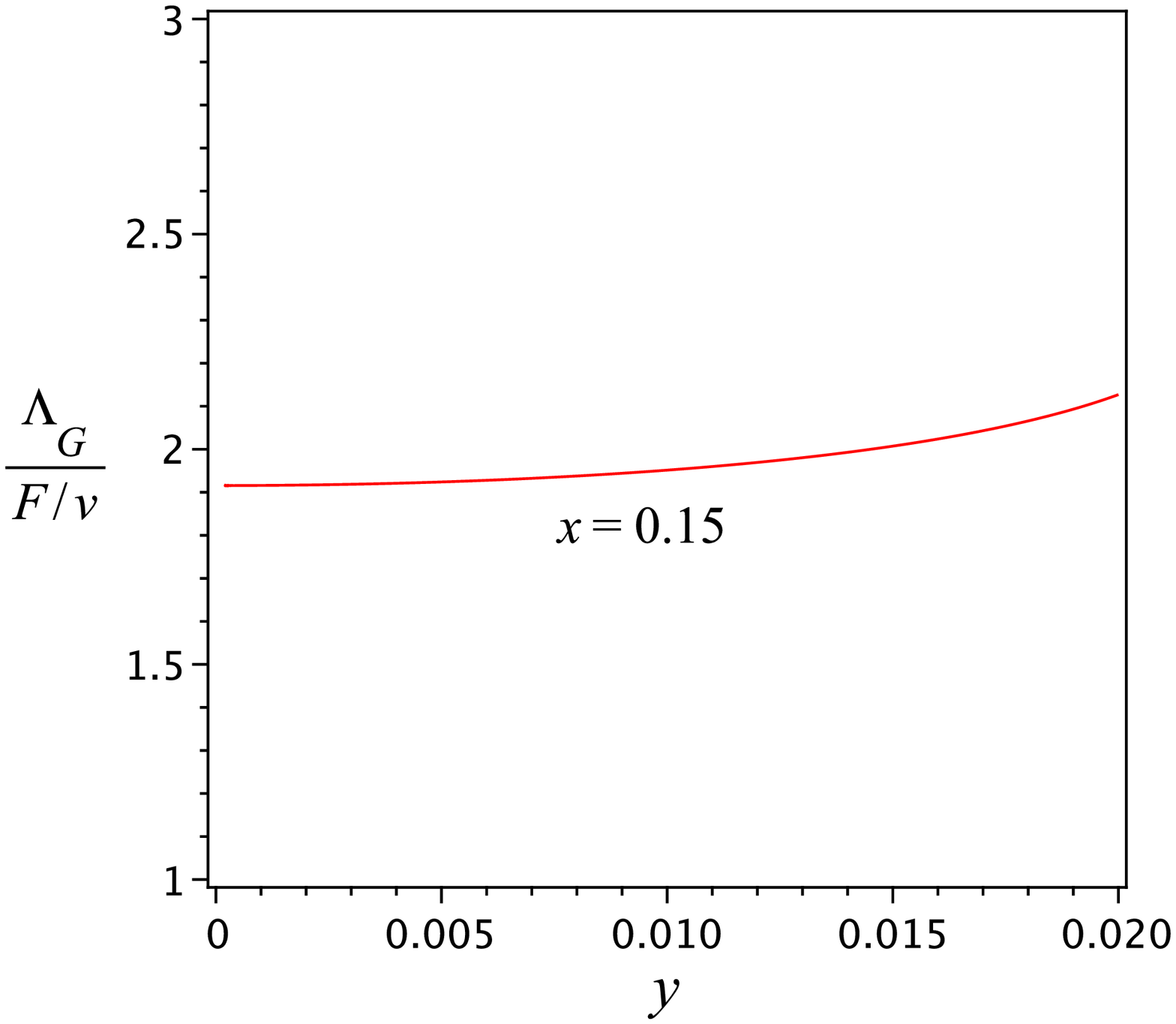}} \scalebox{0.35}{\includegraphics*{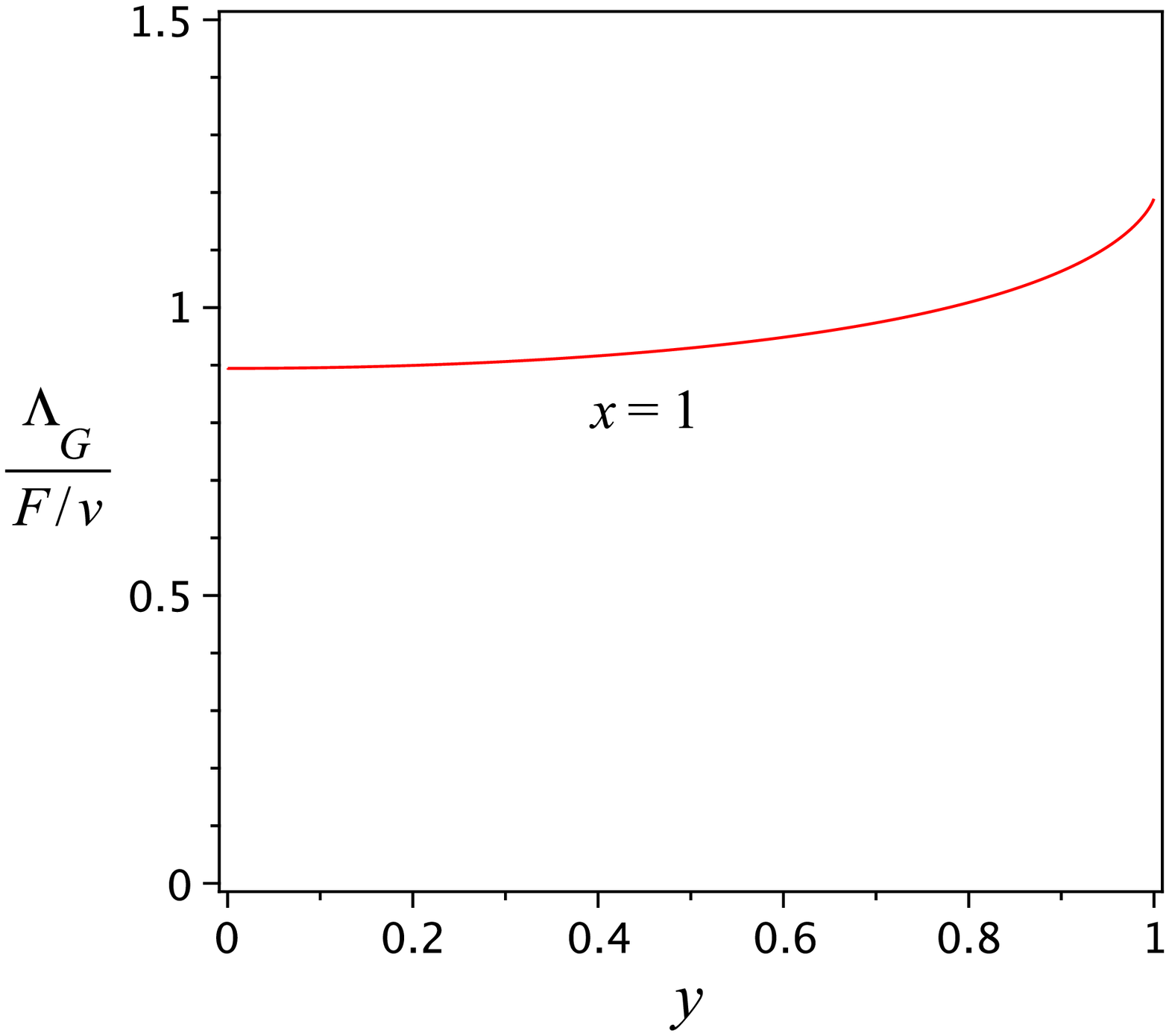}}\\
\scalebox{0.35}{\includegraphics*{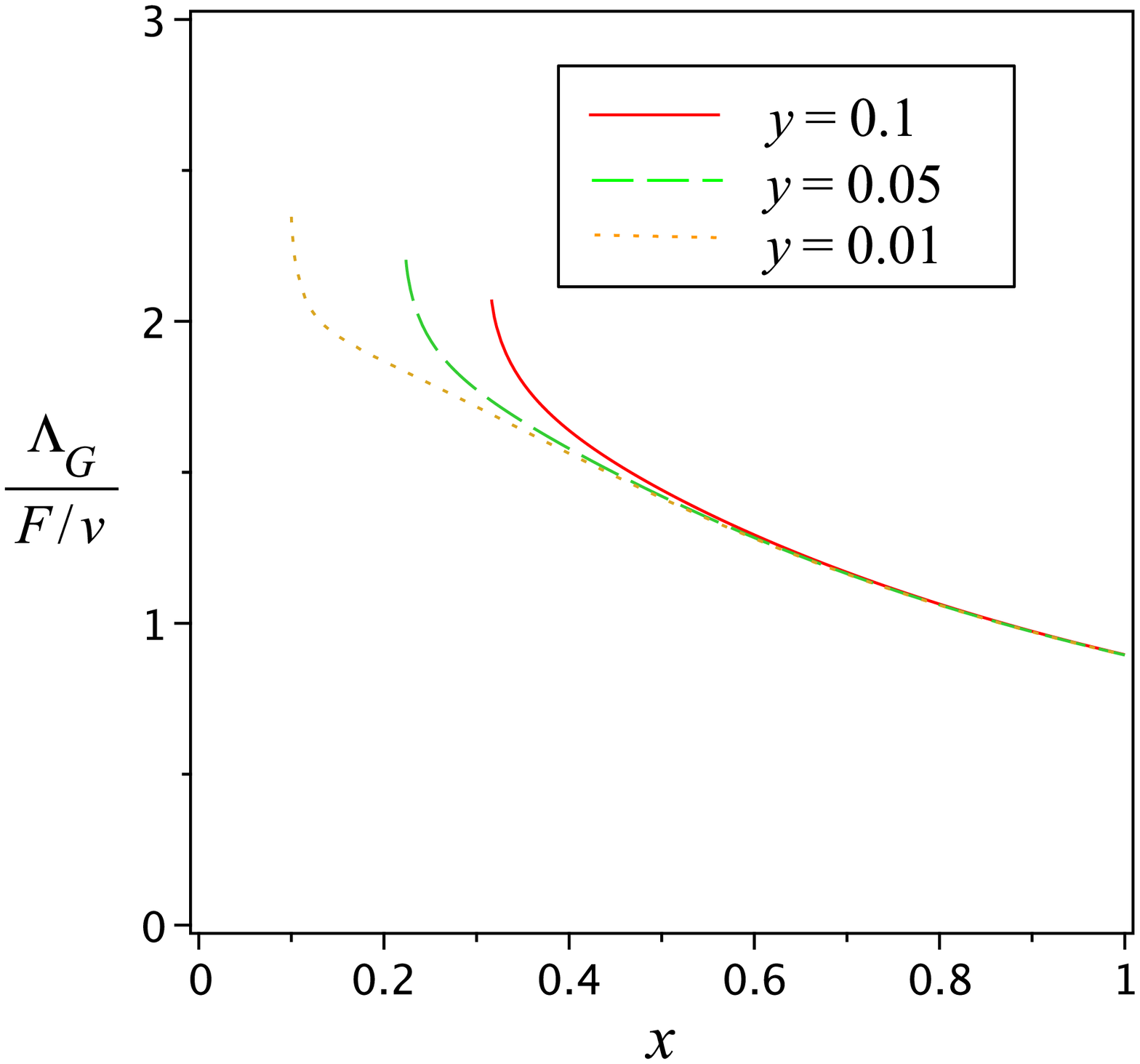}}\caption{The dependences of the dimensionless factor $\Lambda_G/(F/v)$ on $x=m/v$ and $y=F/v^2$.}
\label{fg:gaugino}
\end{figure}

\clearpage

\subsubsection{Sfermion masses}

In ordinary gauge mediation, sfermion masses receive only two-loop contributions from the gauge interactions. In our case, the Yukawa interactions between the matter fields and the Higgs fields give additional one-loop contributions to the sfermion masses, see Fig.\ref{fg:1_loop_sfermion}, and are shown in Table \ref{tb:sfermion}. These formulae are obtained with all Higgs fields circulating in the loops. If instead one integrates out the heavy Higgs fields $H_{\pm 2}$ and does not include the light fields, an ultraviolet divergence will be present  \cite{Poppitz:1996xw}.

\begin{figure}[h]
 \centering
\scalebox{0.8}{\includegraphics*{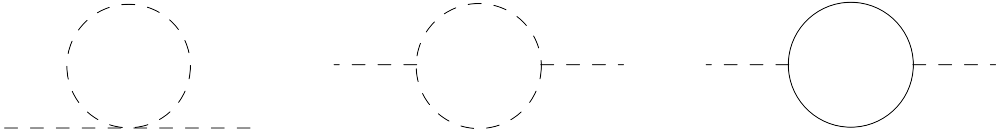}} \caption{One-loop sfermion masses diagrams due to Yukawa interactions. The external lines are the sfermions while the Higgs fields sit inside the loops.}
\label{fg:1_loop_sfermion}
\end{figure}

\begin{table}[h!]
\begin{center}
\begin{tabular}{c|c|c}
\hline
\hline
$\t{Q}_1$ & $(|y_u|^2+|y_d|^2)\Sigma_1$	& $(|y_u|^2+|y_d|^2)\Sigma^\prime_1$ \\
$\t{Q}_2$ & $(|y_u|^2+|y_d|^2)\Sigma_2$	& $(|y_u|^2+|y_d|^2)\Sigma^\prime_2$ \\
\hline
$\t{\bar{u}}_1$ & $2|y_u|^2\Sigma_1$	& $2|y_u|^2\Sigma^\prime_1$\\
$\t{\bar{u}}_2$ & $2|y_u|^2\Sigma_2$	& $2|y_u|^2\Sigma^\prime_2$\\
\hline
$\t{\bar{d}}_1$ & $2|y_d|^2\Sigma_1$	& $2|y_d|^2\Sigma^\prime_1$\\
$\t{\bar{d}}_2$ & $2|y_d|^2\Sigma_2$	& $2|y_d|^2\Sigma^\prime_2$\\
\hline
$\t{L}_1$ & $|y_e|^2\Sigma_1$	& $|y_e|^2\Sigma^\prime_1$\\
$\t{L}_2$ & $|y_e|^2\Sigma_2$	& $|y_e|^2\Sigma^\prime_2$\\
\hline
$\t{\bar{e}}_1$ & $2|y_e|^2\Sigma_1$	& $2|y_e|^2\Sigma^\prime_1$\\
$\t{\bar{e}}_2$ & $2|y_e|^2\Sigma_2$	& $2|y_e|^2\Sigma^\prime_2$\\
\hline
\hline
\end{tabular}
\captionsetup{singlelinecheck=off}
\caption[.]{One-loop sfermion mass squared due to Yukawa interactions. The first column contains contributions from the quartic interactions of Fig.\ref{fg:1_loop_sfermion}, the second from the trilinear interactions. Here the subscript refers to the contribution to a particular family. 
\begin{eqnarray}
(32 \pi^2) \Sigma_1 &=& \cos^2\theta_+ m_{+1}^2\ln m_{+1}^2+\cos^2\theta_- m_{-1}^2\ln m_{-1}^2 -2\cos^2\theta m_{01}^2\ln m_{01}^2 \nonumber \\
& &+ \sin^2\theta_+ m_{+2}^2\ln m_{+2}^2+\sin^2\theta_- m_{-2}^2\ln m_{-2}^2 -2\sin^2\theta m_{02}^2\ln m_{02}^2, \nonumber \\
(32 \pi^2) \Sigma^\prime_1 &=& (m\sin\theta_+)^2\ln m_{+1}^2 + (m\sin\theta_-)^2\ln m_{-1}^2 -2 (m\sin\theta)^2 \ln m_{01}^2 \nonumber \\
& &+ (m\cos\theta_+)^2\ln m_{+2}^2 + (m\cos\theta_-)^2\ln m_{-2}^2 -2 (m\cos\theta)^2 \ln m_{02}^2, \nonumber \\
(32 \pi^2) \Sigma_2 &=& \sin^2\theta_+ m_{+1}^2\ln m_{+1}^2+\sin^2\theta_- m_{-1}^2\ln m_{-1}^2 -2\sin^2\theta m_{01}^2\ln m_{01}^2 \nonumber \\
& &+ \cos^2\theta_+ m_{+2}^2\ln m_{+2}^2+\cos^2\theta_- m_{-2}^2\ln m_{-2}^2 -2\cos^2\theta m_{02}^2\ln m_{02}^2, \nonumber \\
(32 \pi^2) \Sigma^\prime_2 &=& (m\cos\theta_+ +v\sin\theta_+)^2\ln m_{+1}^2 + (m\cos\theta_- +v\sin\theta_-)^2\ln m_{-1}^2 \nonumber
\\
& &-2 (m\cos\theta +v\sin\theta)^2 \ln m_{01}^2 \nonumber \\
& &+ (-m\sin\theta_+ +v\cos\theta_+)^2\ln m_{+2}^2 + (-m\sin\theta_- +v\cos\theta_-)^2\ln m_{-2}^2 \nonumber
\\ 
& &-2 (-m\sin\theta+v\cos\theta)^2 \ln m_{02}^2.\nonumber 
\end{eqnarray}}
\label{tb:sfermion}
\end{center}
\end{table}

We explore same parameter space as for the gaugino masses, with the results:

\begin{enumerate}
\item
For $0 < y \ll x^2 \ll 1$, we find the following expansions for both families,

\begin{eqnarray}
\frac{\Sigma_1+\Sigma^\prime_1}{(F/v)^2} &=& -\frac{1}{3}\left( \frac{y}{x^2} \right)^2-12x^2-8x^2\ln x + \m O\left( \left(\frac{y}{x^2}\right)^4, \left(\frac{y}{x^2}\right)^4x^2, x^4 \ln x  \right), \nonumber \\
\frac{\Sigma_2+\Sigma^\prime_2}{(F/v)^2} &=& 12x^2+8x^2 \ln x + \frac{1}{30}\left( \frac{y}{x^2}\right)^4x^2+ \m O\left(x^4\ln x, \left(\frac{y}{x^2}\right)^6x^2\right).
\end{eqnarray}
The $x^2\ln x$ terms dominate when $x$ approaches 0. The sfermions of the two families have approximately equal masses squared, but they are of opposite signs.

\item
If $0< y \lesssim x^2 \ll 1$, we find 

\begin{eqnarray}
\frac{\Sigma_1+\Sigma^\prime_1}{(F/v)^2} &\sim & \left(x^4+2x^4y+\m O(x^4y^2)\right)\ln \left(1-\frac{y}{x^2}\right), \nonumber \\
\frac{\Sigma_2+\Sigma^\prime_2}{(F/v)^2} &=& (8+6\ln 2)x^2+8x^2\ln x+x^2\left( 1-\frac{y}{x^2}\right)\ln \left(1-\frac{y}{x^2}\right) \nonumber \\ & &+\m O(x^4\ln x, \left(1-\frac{y}{x^2}\right)^2x^2),
\end{eqnarray}
using $(1-y/x^2)$ as one of our expansion parameters. Here the behaviour of the sfermion masses for the first family differs greatly from that of the second.  It can be traced to the logarithmic term $\ln m_{- 1}^2 \sim \ln (1-y/x^2)$, which can give large and negative contributions for as we approach the unphysical point $y = x^2$. (We have assumed $F$ always strictly greater that $m^2$, and the case where they are equal should be taken separately).  

\item
When $0 < y \ll x^2 \lesssim 1$, we have 

\begin{eqnarray} \label{eqn:example}
\frac{\Sigma_1+\Sigma^\prime_1}{(F/v)^2} &= & -\frac{2\sqrt{5}}{125}\left[2\ln \left(\frac{3}{2}+\frac{\sqrt{5}}{2}\right) + 6\sqrt{5}\right ] +\m O(y^2,(1-x)), \nonumber \\
\frac{\Sigma_2+\Sigma^\prime_2}{(F/v)^2} &=& \frac{2\sqrt{5}}{125}\left[2\ln \left(\frac{3}{2}+\frac{\sqrt{5}}{2}\right) + 6\sqrt{5}\right ] +\m O(y^2,(1-x)),
\end{eqnarray}
which show that to leading order, the sfermion masses squared of both families have the exact opposite values. 
\end{enumerate}

These qualitative features can be examined further through numerical analysis. The top left graph of Fig.\ref{fg:sfermion} shows that when $x \ll 1$, the mass squared of the first family switches its sign from positive to negative, and exhibits the divergent behavior as $y$ approaches $x^2$, while the second family depends on $y$ weakly. These observations can be easily explained by combining the above results of the first two cases with $x \ll 1$. The case of $x \lesssim 1$ is given in the top right graph. The mass squared of both families have approximately opposite values, and the mass of the first family becomes large and negative when $y$ is close to one. 
The bottom graph gives the dependence on $x$ for two different choices of $y$. One can see that for the second family, its sfermion mass squared switches signs from negative to positive as $x$ increases, while for the first family it is complicated by the divergence when $x^2$ is near $y$. There exists a small range of $x$ for which the masses of the first family have positive values when $y=0.003$, but if $y=0.01$ they become negative for all allowed values of $x$. This is in agreement with the results that we observed in the top left graph.

\begin{figure}[h]
\centering
\scalebox{0.4}{\includegraphics*{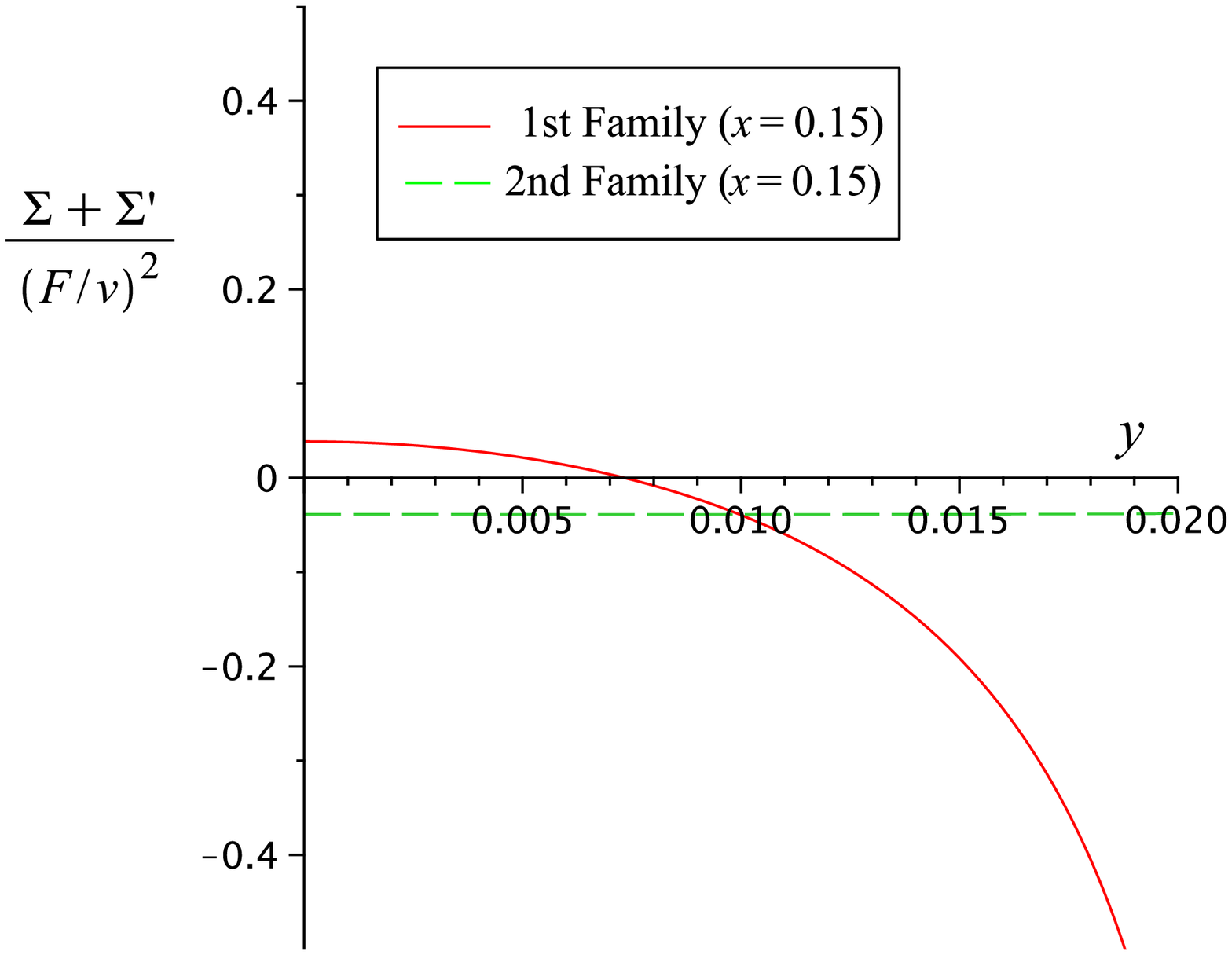}}\scalebox{0.4}{\includegraphics*{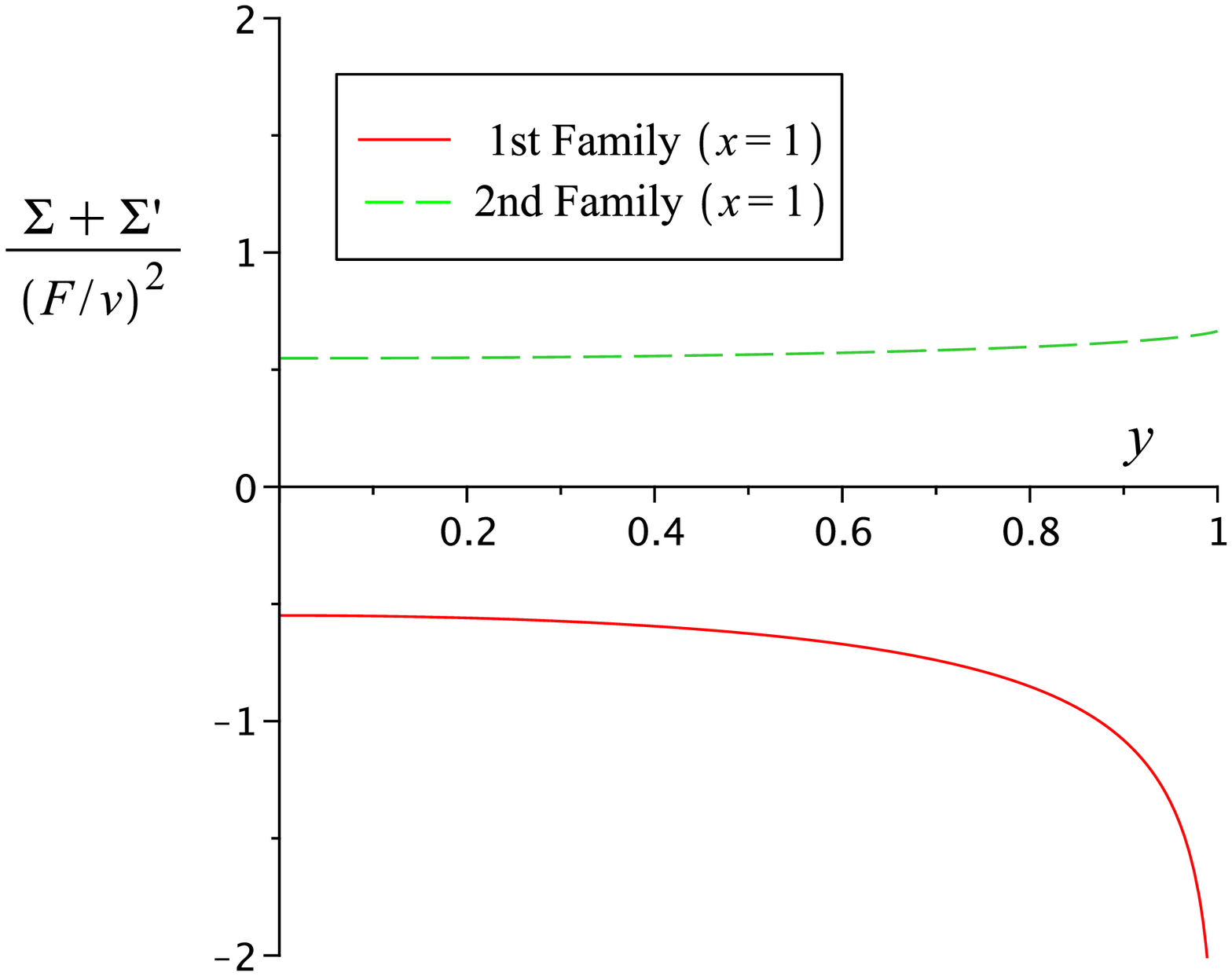}}\\
\scalebox{0.5}{\includegraphics*{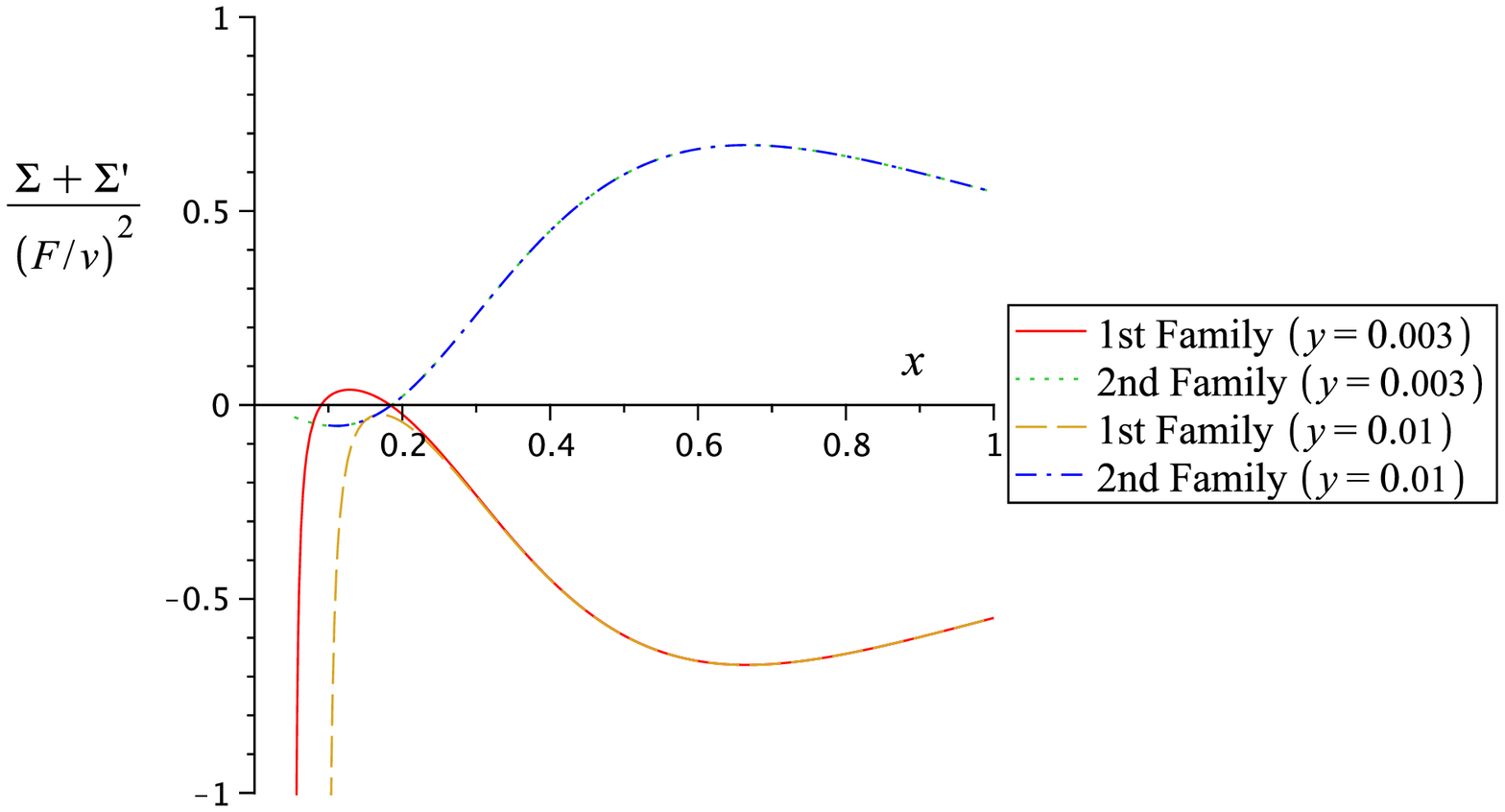}}
\caption{The dependences of the dimensionless factor $\frac{\Sigma+\Sigma^\prime}{(F/v)^2}$ on $x$ and $y$.}
\label{fg:sfermion}
\end{figure}

As seen above, there always exists at least one family of sfermions that picks up negative masses squared, signifying the spontaneous breaking of symmetry. In the slepton sector it leads to two cases:

\begin{itemize}

\item  If only the right handed selectron receives a $vev$, the residual symmetries left over are the gauged $SU(2)$ and a global symmetry $U(1)_{Y+2l}$, $l$ being the global lepton number and $Y$ the hypercharge.

\item If both the right handed selectron and the left handed sneutrino obtain $vev$'s, all gauge symmetries are broken.  We are left with one global $U(1)$ symmetry.

\end{itemize}

A more detailed analysis of such breakings and their resulting spectra are given in Appendix B. None of these produces Standard Model like symmetry breaking. One interesting aspect is that the spectra exhibit different phases for different choices of the parameters of the theory.  Should these one-loop contributions dominate, our model would be a mere curiosity. However, a more realistic scenario emerges when we consider a different vacuum configuration, where family and supersymmetry breakings are misaligned.

\section{Misaligned Vacuum}
\clfnt
\cleqn
In this case, we choose a vacuum where family symmetry breaking is in a different family direction from the supersymmetry breaking:

$$\psi=\begin{pmatrix} 0 \\  1\end{pmatrix},\qquad \psi'=\begin{pmatrix} 1 \\  1\end{pmatrix},$$ 
so that 

\be
{\bf M}=  \begin{pmatrix}
0 & m \\ m & v
\end{pmatrix},\qquad {\bf F}=  F\begin{pmatrix}
1 & 0 \\ 0 & 1
\end{pmatrix},
\ee
displaying the seesaw structure of $\bf M$. $\bf F$ is diagonal and breaks $\m S_3$. In this vacuum $\bf M$ commutes with $\bf F$, resulting in a very distinct structure. 

A single supersymmetric rotation $\m R (\theta)$, independent of supersymmetry breaking, now diagonalizes the Higgs sector. For $m \ll v$, $\theta \approx m/v$ and the fields are separated into decoupled light ($H_{u,d}$) and heavy ($M_{u,d}$) pairs, 

\begin{eqnarray}
\begin{pmatrix} H_{u } \\  M_{u } \end{pmatrix} &=& \m R^t (\theta)\begin{pmatrix} \m H_{u 1} \\ \m H_{u  2} \end{pmatrix}, \qquad \begin{pmatrix} H_{d } \\  M_{d }
\end{pmatrix} = \m R^t (\theta)\begin{pmatrix} \m H_{d 1} \\ \m H_{d  2} \end{pmatrix} , 
\end{eqnarray} 
leading to the superpotential 

\begin{eqnarray}
W_M = m_{\t H} H_uH_d + m_{\t M} M_uM_d+\theta^2 FH_uH_d +\theta^2FM_uM_d,
\end{eqnarray}
where $m_{\t H} \approx \frac{m^2}{v}$ and $m_{\t M} \approx v$ are the masses of the light and heavy Higgsino pairs, respectively. From,

\be
\det (\B M^\dagger \B M \pm \B F) = (m^2 \pm F)^2  \pm F v^2,
\ee
we find the condition for a negative eigenvalue, 

\begin{equation}
\frac{F}{v^2} > \left( \left(\frac{m}{v}\right)^2- \frac{F}{v^2} \right)^2.
\end{equation}
As in the aligned vacuum we will have negative eigenvalues for a certain range of the parameters. Spontaneous breaking of symmetry will occur at tree-level as in the previous case (see Appendix A), and we will not pursue it further.  

The Yukawa superpotential, when written in terms of the light and heavy fields, becomes

\begin{equation} \label{eq:yukawa}
W_Y = Q \B Y_u \bar{u} H_u + Q \B Y_u^\prime \bar{u} M_u + Q \B Y_d \bar{d} H_d + Q \B Y_d^\prime \bar{d} M_d +  L \B Y_e \bar{e} H_d + L \B Y_e^\prime \bar{e} M_d,
\end{equation}
where 

\begin{eqnarray} \label{eqn:yukawamatrix}
\B Y_u &=& y_u \begin{pmatrix} \cos \theta & 0 \\ 0 & \sin \theta  \end{pmatrix} \equiv \begin{pmatrix}
y_{u1} & 0 \\ 0 & y_{u2} \end{pmatrix}  \approx  y_u
\begin{pmatrix} 1 & 0 \\ 0 & \frac{m}{v} \end{pmatrix}   ,\\ \nonumber
\B Y_{u}^\prime &=& y_{u} \begin{pmatrix} -\sin \theta & 0 \\ 0 & \cos \theta \end{pmatrix} \equiv \begin{pmatrix}
y_{u1}^\prime & 0 \\ 0 & y_{u2}^\prime \end{pmatrix} \approx y_{u}
\begin{pmatrix} -\frac{m}{v} & 0 \\ 0 & 1 \end{pmatrix},
\end{eqnarray}
are family space matrices, and we have distinguished the couplings to the heavy messengers with a prime. The Yukawa matrices for the down quark and lepton sector are identical to those of Eq.(\ref{eqn:yukawamatrix}) with the replacement of $y_u$ with $y_d$ and $y_e$.  The matrices $\B Y$ and $\B Y^\prime$ have an inverse hierarchy that may be traced to the family group structure. 

With the model specified at tree-level, we next turn to the radiative corrections.

\subsection{Radiative Corrections}

The fact that $\B M$ and $\B F$ commute simplifies the radiative structure. As we remarked earlier, at tree level the Higgs fields split into light and heavy set, independent of supersymmetry breaking.  At the loop level, the ultraviolet divergences are not present.  

We can integrate out the heavy fields $M_{u,d}$ to generate soft terms and act as messengers of supersymmetry breaking. This case most resembles gauge mediation, with the messengers allowed to interact with the matter fields. The soft terms are calculated at the mass scale of the heavy Higgs fields; for $F/v^2 \ll 1$, this is approximately $v$. 

For the bino and winos, Eq.(\ref{eqn:gauginomass}) simplifies to the usual one-loop gauge mediated result,

\begin{equation}
M_1 = \frac{g^{\prime 2}}{16 \pi^2} \frac{F}{v}, \qquad
M_2 = \frac{g^{2}}{16 \pi^2} \frac{F}{v}. 
\end{equation}

Since $\B M$ and $\B F$ commute, the soft sfermion masses are qualitatively different from the previous case. One-loop sfermion masses generated by Yukawa interactions vanish to leading order in SUSY breaking due to an accidental cancellation, so that their leading contributions occurs at two-loops.

This cancellation occurs in a wide variety of models as noted by \cite{Dine:1996xk,Dvali:1996cu}. To understand the generic features which lead to such a cancellation, consider a superpotential of the form

\begin{eqnarray}
W = y_{\psi} \lambda_{ijk} \psi_i \bar{\psi}_j \m H_{k} + \B M_{ij}\m H_{u i}\m H_{d j} + \theta^2  \B F_{ij} \m H_{u i} \m H_{d j},
\end{eqnarray}
where $\psi_i$ and $\m H_k$ are $SU(2)$ doublets, $\bar{\psi}_j$ is an $SU(2)$ singlet, $y_\psi$ is a Yukawa coupling, and $\lambda_{ijk}$ are CG coefficients. The one-loop sfermion masses squared are 

\begin{equation}\label{eq:oneloopsfermion}
m_{\t{\bar{\psi}}_i}^2 =  2 m_{\t{\psi}_i}^2 = 2 |y_\psi|^2 (\Sigma_i+\Sigma^\prime_i), 
\end{equation}
where,
\begin{eqnarray} \label{eqn:sigmas}
(32 \pi^2 )\Sigma_i &=& \sum_{j,r} \sum_{\pm} [  \lambda_i \m R(\theta) ]_{jr}^2 \left(m_{0r}^2 \pm  f_{r} F\right) \ln (\left(m_{0r}^2 \pm  f_{r} F\right))-2 \sum_{j,r} [  \lambda_i \m R(\theta) ]_{jr}^2 m_{0 r}^2 \ln (m_{0 r}^2) \nonumber \\
(32 \pi^2 ) \Sigma^\prime_i &=& \sum_{j,r} \sum_{\pm} [ \lambda_i \B M^t \m R(\theta) ]_{jr}^2 \ln (\left(m_{0r}^2 \pm  f_{r} F\right)) -2 \sum_{j,r} [ \lambda_i \B M^t \m R(\theta) ]_{jr}^2 \ln (m_{0 r}^2), 
\end{eqnarray}
where $(\lambda_i)_{jk} \equiv \lambda_{ijk}$, and $f_r F$ is the $r$-th eigenvalue of $\B F$. The extra factor in $\t{\bar{\psi}}$ comes from $SU(2)$.

For any $\lambda_i$,

\begin{eqnarray} \label{eqn:proof1}
\sum_j [\lambda_i \B M^t \m R(\theta) ]_{jr}^2 = \sum_{j}  [ \lambda_i \m R(\theta) ]_{jr}^2 m_{0r}^2.
\end{eqnarray}
This equality holds, as can be seen by writing it in matrix form,

\begin{eqnarray}
\Big ( \m R(\theta)^t \B M \lambda_i^t \lambda_i \B M^t \m R(\theta)\Big )_{rr} = \Big ( \m R(\theta)^t \lambda_i^t \lambda_i  \m R(\theta)\Big )_{rr} \Big ( \B M_0^2\Big )_{rr},
\end{eqnarray}
since $R(\theta)^t \B M R(\theta) =\B M_0$. Now, the $F$ dependent part of Eq.(\ref{eq:oneloopsfermion}) is of the form 

\begin{eqnarray}
\label{eq:vanishing_form}
\sum_{\pm}[(a\pm bz)\ln(a\pm bz)+ a \ln(a\pm bz)], \qquad a,b>0,
\end{eqnarray}
whose expansion for $bz/a \ll 1$ (equivalent to the constraint $f_r F/m_{0r}^2 \ll 1$ ) does not contain a term of order $z^2$. The one-loop Yukawa contributions to the sfermion masses squared vanishes at leading order of SUSY breaking - there are no $F^2$ terms in their expansions. 

The one-loop contributions are now of order $F^4$, and the two-loop contributions to the sfermion masses, of order $F^2$, are dominant for a range of parameters. They include the usual gauge mediated results which go like $g^4$ \cite{Giudice:1998bp}, and new two-loop Yukawa/gauge contributions which go like $g^2 y^2$ and $y^4$.  Using wave function renormalization techniques \cite{Giudice:1997ni}, we find \footnote{Our results agree with those in \cite{Abdullah:2012tq}, but disagree with those in \cite{Evans:2012hg}.}

\begin{eqnarray} \label{eqn:gymixed}
\delta m_{\t Q_i}^2 &=& \frac{1}{256 \pi ^4} \Bigg[ y_{u i}^{\prime 2} \left( 3 y_{u i}^{\prime 2} + 3 \displaystyle\sum\limits_{j} y_{u j}^{\prime 2} + y_{d i}^{\prime 2} - 2 \left( \frac{8}{3} g_s^2 + \frac{3}{2} g^2 + \frac{13}{18} g^{\prime 2} \right) \right)  \\ \nonumber
&+& y_{d i}^{\prime 2} \left( 3 y_{d i}^{\prime 2} +  \displaystyle\sum\limits_{j} ( 3 y_{d j}^{\prime 2} +y_{e j}^{\prime 2} ) + y_{u i}^{\prime 2} - 2 \left( \frac{8}{3} g_s^2 + \frac{3}{2} g^2 + \frac{7}{18} g^{\prime 2} \right) \right) \Bigg]\left|\frac{F}{v}\right|^2 ,\\ \nonumber
\delta m_{\t{\bar{u}}_i}^2 &=& \frac{1}{128 \pi ^4} \Bigg[  y_{u i}^{\prime 2} \left( 3 y_{u i}^{\prime 2} + 3 \displaystyle\sum\limits_{j} y_{u j}^{\prime 2} + y_{d i}^{\prime 2} +y_{d i}^2 - \left( \frac{16}{3} g_s^2 + 3 g^2 + \frac{13}{9} g^{\prime 2} \right) \right) - y_{u i}^2 y_{d i}^{\prime 2} \Bigg] \left|\frac{F}{v}\right|^2 ,\\ \nonumber
\delta m_{\t{\bar{d}}_i}^2 &=& \frac{1}{128 \pi ^4} \Bigg[  y_{d i}^{\prime 2} \left( 3 y_{d i}^{\prime 2} +  \displaystyle\sum\limits_{j} (3 y_{d j}^{\prime 2} + y_{e j}^{\prime 2}) + y_{u i}^{\prime 2} +y_{u i}^2 - \left( \frac{16}{3} g_s^2 + 3 g^2 + \frac{7}{9} g^{\prime 2} \right) \right) - y_{d i}^2 y_{u i}^{\prime 2} \Bigg] \left|\frac{F}{v}\right|^2, \\ \nonumber
\delta m_{\t{L}_i}^2 &=&  \frac{1}{256 \pi ^4} \Bigg[ y_{e i}^{\prime 2} \left( 3 y_{e i}^{\prime 2} +  \displaystyle\sum\limits_{j} (3 y_{d j}^{\prime 2} + y_{e j}^{\prime 2}) -  3 \left(  g^2 +  g^{\prime 2} \right) \right) \Bigg] \left|\frac{F}{v}\right|^2 ,\\ \nonumber 
\delta m_{\t{\bar{e}}_i}^2 &=& 2 \Delta m_{\t{L}_i}^2,  \nonumber \\
\delta m_{H_u}^2 &=& - \frac{1}{256 \pi ^4} \Bigg[ \displaystyle\sum\limits_{j} 3 y_{u j}^2 \left( 3 y_{u j}^{\prime 2} + y_{d j}^{\prime 2} \right)  \Bigg] \left|\frac{F}{v}\right|^2, \nonumber \\
\delta m_{H_d}^2 &=&  - \frac{1}{256 \pi ^4} \Bigg[ \displaystyle\sum\limits_{j} 3 y_{d j}^2 \left( 3 y_{d j}^{\prime 2} + y_{u j}^{\prime 2} \right) + \displaystyle\sum\limits_{j} 3 y_{e j}^2 y_{e j}^{\prime 2}  \Bigg] \left|\frac{F}{v}\right|^2 \nonumber,
\end{eqnarray}
where $g_s$ is the $SU(3)$ gauge coupling.  They are supplemented by one-loop A terms,

\begin{eqnarray}
a_{u i} &=& - \frac{1}{16 \pi^2} y_{u i} \left( 3 y_{u i}^{\prime 2} + y_{d i}^{\prime 2} \right) \left|\frac{F}{v}\right| ,\\ \nonumber
a_{d i} &=& - \frac{1}{16 \pi^2} y_{d i} \left( 3 y_{d i}^{\prime 2} + y_{u i}^{\prime 2} \right) \left|\frac{F}{v}\right| ,\\ \nonumber
a_{e i} &=& - \frac{3}{16 \pi^2} y_{e i} y_{e i}^{\prime 2} \left|\frac{F}{v}\right| ,
\end{eqnarray} 
which differ from the results of normal gauge mediation where $A$ terms are generated at two-loops. \footnote{Without an extra loop suppression, larger $A$ terms at the boundary are expected. However, the structure of Eq.(\ref{eqn:yukawamatrix}) gives additional suppression by factors of $\sin\theta$ or $\sin^2\theta$ relative to the other soft mass parameters. }

These soft terms serve as boundary conditions for the renormalization group at $v$, the mass of the heavy fields. 

The gaugino masses have the same structure as in normal gauge mediation, but the soft sfermion masses now exhibit a family dependent hierarchy at the messenger scale, generated by the two-loop Yukawa/gauge contributions of Eq.(\ref{eqn:gymixed}). 

Their qualitative features depend on the relative magnitudes of the gauge and Yukawa couplings, as we can see by Eq.(\ref{eqn:gymixed}). 

\begin{itemize}
\item
For $g^2 \gg y^{\prime 2}$, the full two-loop contributions go like $g^4$, with a negative correction which goes like $y^{\prime 2} g^2$. The sfermion masses with the larger $y^\prime$ couplings will be lighter. 

\item
When $y^{\prime 2} \gg g^2$, the full two-loop soft masses are dominated by the positive $y^{\prime 4}$ terms, and now the family with the larger $y^{\prime}$ couplings will be heavier. 

\item 
The intermediate case, where the Yukawa and gauge couplings are comparable, is more complicated as all terms are of the same order. Here the order of the masses may not necessarily be ordered by $y^\prime$.  Furthermore, one must take care to ensure that none of these masses squared becomes negative.
\end{itemize}

In our model, the family hierarchy of the sfermion masses at the boundary is purely determined by the \emph{primed} Yukawa couplings $\B Y^\prime$, which have the hierarchy $y_1^\prime \ll y_2^\prime$.\footnote{The apparent $\B Y$ dependence of the masses for the right handed up and down squarks cancels, since in this model $(y_{u i}^{\prime 2}  y_{d i}^{2} - y_{u i}^{2}  y_{d i}^{\prime 2} )$ vanishes.} This stems from both the CG coefficients and the rotation matrix $\m R(\theta)$.  $\m S_3$ couples the first family quarks and leptons only to $\m H_{u 1}$ and the second family only to $\m H_{u2}$. When the rotation angle is small, the largest component of $H_u$ is $\m H_{u1}$, while for $M_u$ it is $\m H_{u2}$, as  reflected in Eq.(\ref{eqn:yukawamatrix}).  Qualitatively, the sfermion masses of the first family will have larger (smaller) masses than those of the second when $g^2 \gg y^{\prime 2}$ ($g^2 \ll y^{\prime 2})$. Contrast this with the usual paradigm of generating flavor blind (tasteless) soft parameters at the boundary to ensure nearly  degenerate squark masses. 

However, the physical masses are determined by these couplings at much lower energies, related to the boundary values through RG running.  What happens to this hierarchy as we run to low energy? 

To explore this question qualitatively, we use the one-loop RG equations of the MSSM, which coincide with those of our model when modified to include only two families.  Generically, the one-loop RG equations for the sfermion soft masses are of the form 

\begin{equation}
\frac{d}{dt} m_{\t f}^2 \sim c_y y^2 - c_g g^2,
\end{equation} 
where $c_y$ and $c_g$ are positive functions of the soft masses.\footnote{There are also positive contributions from the $A$ terms that go like $a^2$. We have not shown them here because the RG running behavior of the $A$ terms mainly depend on the Yukawa couplings $y$, so that their impact on the family splitting is similar to the above $y^2$ terms. } Gauge interactions are family independent, and the family structure of the running is dictated by the Yukawa interactions. Positive contributions from the Yukawa interactions decrease the soft masses from their boundary values as we run to the infrared, while the negative gauge contributions have the opposite effect.  The overall behavior of the running will of course depend on the relative magnitudes of the Yukawa and gauge contributions. Since the RG equations for the mass difference between the first and second family sfermions is of the form

\begin{equation}
\frac{d}{dt} (m_{\t f 1}^2 - m_{\t f 2}^2) \sim (y_1^2 - y_2^2),
\end{equation}
we may distinguish two qualitatively distinct cases. 

\begin{itemize}
\item
In the first, the family with the larger Yukawa coupling is also the family with the larger boundary value mass squared. In this case, the hierarchy at the large scale will be attenuated in the infrared by RG running. 
\item
For the second case, the family with the larger Yukawa coupling is lighter at the boundary.  Now, the initial hierarchy will instead grow as we approach the infrared.
\end{itemize}

When Yukawa interactions are allowed to participate in the breaking of supersymmetry, the family structure of the soft masses at the boundary and their RG running are correlated, depending on how the matter fields couple to the light and heavy Higgs fields.  These couplings, as in this model, may be related by a family symmetry.  There may be models such that a splitting at the boundary is erased in the infrared.

This model provides an example of this mechanism.  At $v$, the large Yukawa couplings of the second family to the heavy messengers suppresses their masses with respect to those of the first family.  During RG running, their small Yukawa couplings to the light Higgs fields tend to decrease this initial splitting (qualitatively like the first case discussed above).  

However, this is not the full story.  The physical criterion we need to satisfy is a near degeneracy of the physical masses. These physical masses are defined as the values of the running masses when these values are commensurate with the scale. This gives an additional constraint, and whether this will be possible will depend critically on both the scale of the boundary and the initial splitting of the soft masses. 

Moreover, if the running soft masses were to meet at one point and then diverge, arranging this point to be at the scale of their physical masses may be considered extreme fine tuning. If instead the masses become nearly degenerate over a large range of scales, this focusing may well be a generic feature of the model.    

\subsection{Numerical RG Study}

We verify numerically the qualitative features using the one-loop RG equations without threshold corrections and the boundary conditions just discussed. For the purpose of RG running, we add gluino masses by hand at the boundary. They are chosen to be the usual one-loop gauge mediated results, $M_3 = (g_s^2/16 \pi^2 ) F/v$. 

The running of the right handed down squark mass is shown in Fig.\ref{fg:Urunning}. It displays the focusing behavior most clearly due to the large up type Yukawa couplings. For illustrative purposes we have shown the results for two sets of parameters. In both cases we see that the masses stay nearly degenerate for an appreciable range, however, for the first set of parameters the scale at which they meet is orders of magnitude larger than their values.

To ensure that $m$ and $F$ are chosen such that the lightest Higgs mass eigenstate be positive, we trade our free parameters $m$ and $F$ for $m_{-1}^2$ and $m_{\t H}$, the mass of the lightest Higgsino, 

\begin{eqnarray}
m \simeq \sqrt{m_{\t H} v}, \qquad F = m_{\t H}^2 - m_{-1}^2. 
\end{eqnarray}   
For the first set of parameters we choose $v = 10^8$ $GeV$, $m_{-1} = 10^2$ $GeV$, and $m_{\t H} = 10^3$ $GeV$. For the gauge and Yukawa couplings we consider the set of values $y_u = 0.96$, $y_d = y_e = 0.1$ , $g^\prime = 0.39$, $g = 0.65$, and $g_s = 1.02$. The different magnitudes of these couplings for each sector cover the qualitative ranges discussed previously.

\begin{figure}[!h]
 \centering
\scalebox{0.4}{\includegraphics*{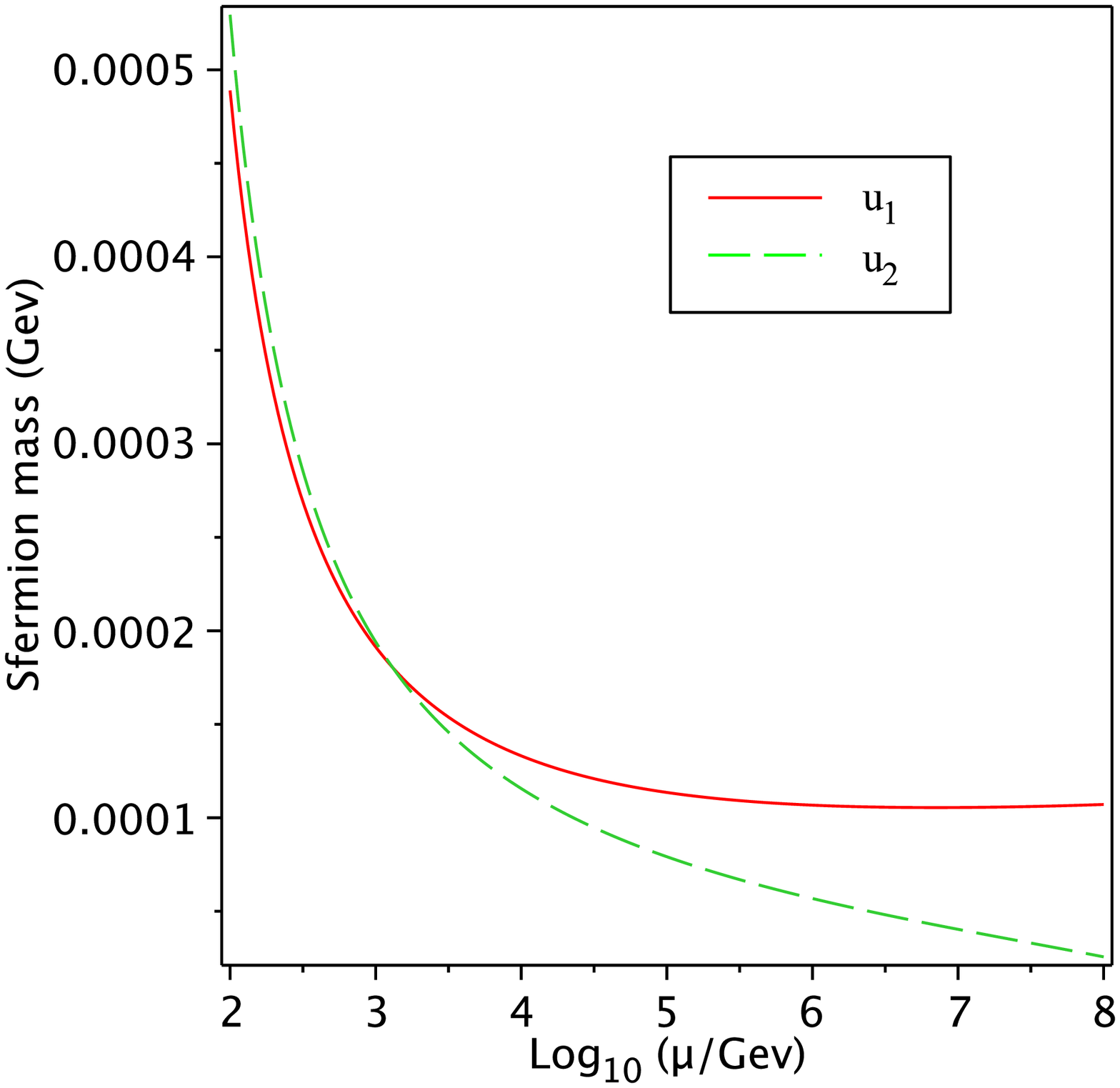}} \scalebox{0.4}{\includegraphics*{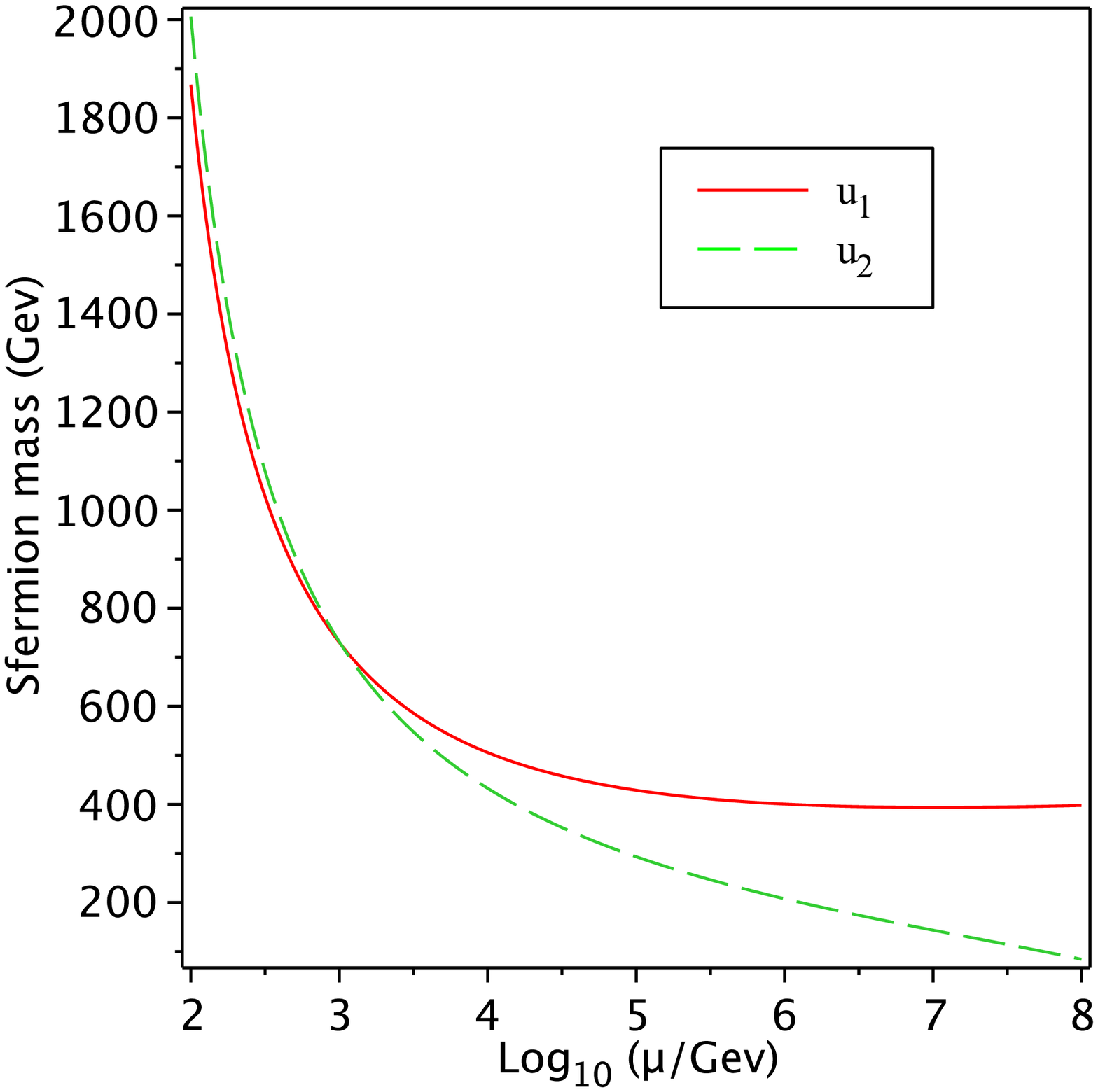}}
\caption{RG running of $m_{\t{\bar{u}}_i}$. In the first figure, $v=10^8$ $GeV$, $m_{-1} = 10^2$ $GeV$, $m_{\t H} = 10^3$ $GeV$,  $y_u = 0.96$, $y_d = y_e = 0.1$ , $g^\prime = 0.39$, $g = 0.65$, and $g_s = 1.02$.  In the second, the other parameters are kept the same except $m_{\t H} = 2 \times 10^6$ $GeV$.}
\label{fg:Urunning}
\end{figure}

In the second trial run we set $m_{\t H} = 2 \times 10^6$ $GeV$ to increase the value of $F/v$, raising the overall scale of the soft masses, while the other parameters are kept the same.\footnote{The fact that we must increase the mass of the light Higgsinos and Higgs fields will introduce some error, as technically one should integrate them out during the running when passing the scale of their masses.}
  
Although this provides a proof of concept, the overall scale of these masses is unrealistic in this model, as the light higgsino mass $m_{\t H}$ is much larger than the soft masses. The soft masses at the boundary are generically of the order

\begin{equation}
m_{soft}^2 \sim \frac{\m O(1)}{16 \pi^2}  \left(\frac{F}{v}\right)^2.
\end{equation} 
However, the requirement that the lightest of the Higgs fields have a positive mass squared puts a constraint on $F$, $F < m_{\t H}^2$, so that 

\begin{equation}
m_{soft}^2 \lesssim \frac{\m O(1)}{16 \pi^2} \left(\frac{m_{\t H}^2}{v}\right)^2 \ll m_{\t H}^2.
\end{equation}
However, an inspection of Fig.\ref{fg:Urunning} shows that the focusing feature is quite independent of this scale problem, and may be solved separately. It may be alleviated in a more complete model that includes various sources of supersymmetry breaking, or where a parameter like $m$ is not present. We believe its solution is a model dependent question.

The running of the other soft masses can be found in Appendix C.  When the Yukawa couplings are small, as for the down squark and slepton sectors, the focusing mechanism is not observed.

As noted in the previous section, the splitting at the boundary and behavior during RG running are correlated through the relations satisfied by $\B Y$ and $\B Y^\prime$. In another model that exhibits this focusing mechanism, the relationship between mass splitting and RG running may be quite different.  

However, the Yukawa couplings and RG running of the MSSM are known and may provide model independent, bottom-up guidelines for model building. Given degenerate physical masses, how large is their splitting at the high energy boundary?

\begin{figure}[!h]
 \centering
\scalebox{0.38}{\includegraphics*{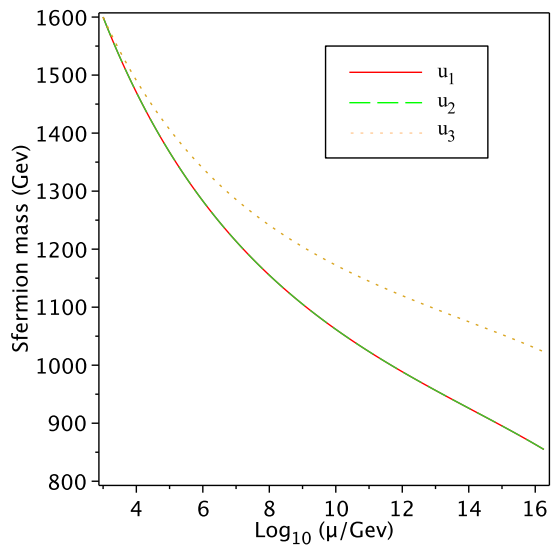}}
\caption{RG running of $m_{\t{\bar{u}}_i}$ with degenerate $TeV$ masses in the MSSM with $\tan \beta = 10$. The soft masses of the first two family members are indistinguishable because of their small differences in Yukawa couplings.}
\label{fg:Urunning_MSSM}
\end{figure}

To answer this, we use the one-loop RG equations of the MSSM and impose the boundary condition that the soft sfermion masses be degenerate in the infrared. The results for the right handed up squarks are shown in Fig.\ref{fg:Urunning_MSSM}.  We see that the soft masses of the first two family members are indistinguishable because of their small differences in Yukawa couplings, but a large splitting exists between the second and third generation. For model building purposes, figures such as this may be used to put constraints on the ultra violet theory. For example, if it was a Grand Unified Theory and no intermediate particles were present that would effect the RG running, Fig.\ref{fg:Urunning_MSSM} would tell us that the theory at the GUT scale should produce a splitting of $\sim 200$ $GeV$ for the right handed squarks.  Given the phenomenologically allowed splitting at low energy, the splitting at a given high energy scale may be inferred.  

We do not touch here the question of electroweak symmetry breaking, although an inspection of Appendix C shows that one of the Higgs masses does indeed run negative.

\section{$\m Z_2$ Invariant Vacuum }
\clfnt
\cleqn
Finally we consider the simplest vacuum structure where 

$$\psi= \psi'=\begin{pmatrix} 1 \\  1\end{pmatrix},$$
which breaks down $\m S_3$ to $\m Z_2$. We lose the seesaw as

\be
{\bf M}=  \begin{pmatrix}
v & m \\ m & v
\end{pmatrix},\qquad {\bf F}=  F\begin{pmatrix}
1 & 0 \\ 0 & 1
\end{pmatrix}.
\ee
In this case, the mixing angles simplify considerably:

\begin{equation}
\tan \theta_\pm = \tan \theta = -45^\circ,
\end{equation}
and the eigenvalues are given by 

\begin{eqnarray}
m^2_{\pm 1} &=& (v-m)^2 \pm F, \\ \nonumber
m^2_{\pm 2} &=& (v+m)^2 \pm F . \\
\end{eqnarray}
There is no see-saw, but if $m$ is sufficiently close to $v$  a hierarchy is still possible.  The condition for a negative eigenvalue, given that $F \ll v^2$, is  

\begin{equation}
F > (v-m)^2,
\end{equation}
which will depend on how close to $v$ we choose $m$. 

Again, the case of a negative eigenvalue follows closely the results for the aligned vacuum.  When all masses are positive, the general features for this model follow that of case misaligned vacuum.  Here $\B M$ and $\B F$ commute, so again the two-loop contributions are dominant.  The results of RG running also follow the general trends of that for the aligned vacuum, but do not exhibit the interesting focusing feature, as the residual $\m Z_2$ symmetry gives degenerate soft terms which simply run together. Again, demanding that one of the Higgs fields have a mass of the weak scale constrains the overall scale of the soft masses.  Since no new interesting feature emerges in this vacua, we leave it here and turn to a summary of our results.

\section{Summary and Conclusions}

We set out to study the marriage of family and supersymmetry breakings in a model where the Higgs sector is extended to include family partners which serve as messengers of supersymmetry breaking. It is believed that such family dependent supersymmetry breaking would be ruled out by FCNC constraints. We have been able to show qualitative features that suggest that such models may not be ruled outright.  

Their phenomenology is determined by the relative alignment of family and supersymmetry breakings.  We organize these vacuum configurations in terms of the supersymmetric mass matrix of the Higgs fields $\B M$, and the supersymmetry breaking matrix $\B F$. In all cases, for some range of parameters we have spontaneous breaking of symmetry at tree level and negative soft parameters.  When there is no breaking of symmetry at tree level, the phenomenology is mainly driven by radiative corrections due to the Yukawa interactions not present in standard gauge mediation. 

When $\B M$ and $\B F$ do not commute, as was the case for the aligned vacuum,  the soft terms are dominated by one-loop corrections.  Some sfermion masses squared can be negative, leading to phenomenologically disastrous patterns of symmetry breaking. In addition, an ultraviolet divergence is present if one integrates out the heavier fields. 

If $\B M$ and $\B F$ do commute, as for the misaligned and $\m Z_2$ invariant vacua, the one-loop contributions vanish at leading order in supersymmetry breaking.  The dominant contribution to the soft terms now comes from two-loops, and no divergence is present if the heavier Higgs fields (when a large hierarchy in the Higgs sector exists) are integrated out. 

These two-loop contributions generically produce family dependent sfermion masses at the messenger scale.\footnote{An exception was the $\m Z_2$ invariant vacuum.} However, there exists a focusing mechanism by which such hierarchies can in principle be erased by RG running. In our model, this was due to a \emph{correlation} between the mass splittings of the sfermions at the boundary, controlled by the Yukawa couplings of the sfermions to the heavy Higgs fields, and their RG running, controlled by their Yukawa couplings to the light Higgs fields. We were able to identify the generic features needed for such a mechanism to take place, and verify the effect numerically in our toy model, although the simplicity of our model led to a problem with the scale of soft masses generated.  

This brings the exciting possibility of new tasteful supersymmetric model building. In future work, we would like to consider more realistic models that mimic the general features of this focusing mechanism.

A generalization of this model would be to choose a larger family group that includes triplet representations and is capable of producing a realistic spectrum.  Although the number of parameters will increase, it remains to be seen whether the important features of this toy model can be reproduced: a hierarchy between the Higgs family partners, supersymmetric mass and supersymmetry breaking matrices that commute, and a hierarchy in Yukawa couplings of the matter fields to the light and heavy Higgs fields.  

In addition, the scale of soft masses needs to be addressed. It is a question that is intimately connected with the question of electroweak symmetry breaking, which will also need to be worked out in detail.  In this regard, these models of flavored supersymmetry breaking tie in the old ``$\mu$ problem'' of the MSSM. The initial question of whether a family symmetry could forbid or suppress the $\mu$ term was not fully addressed in the $\m S_3$ model, as the family group allowed for a ``$\mu$ term'', $m$. A small effective $\mu$ was instead possible because of a see-saw structure in $\B M$.  

The model contained no color triplet messengers.  In a grand unified model, we will have family dependent color triplet messengers which will give mass to the gluinos. Their role, and the question of their contributions to proton decay are left for future work. 

Our model, incomplete as it is, has shown that the marriage of family and supersymmetry deserves further attention. Of special interest is the focusing mechanism, which depends crucially on the interplay between boundary conditions and renormalization group running. With some idea of the pros and cons, we would like to continue to study this idea in a more complete model.

\section{Acknowledgements}

We thank Jesus Escobar for his useful discussions at the early stages of this work.  The authors are also very grateful to Graham Ross and AseshKrishna Datta for their useful comments and insights. PR thanks the Aspen Center for Physics for its hospitality, where part of this work was done. This research is partially supported by the Department of Energy Grant No. DE-FG02-97ER41029.

\newpage

\appendix
\appendixpage

\section{Detailed Spectrum for $m^2 < F$}

We present here the details of the mass spectrum for the aligned vacuum configuration when $m^2 < F$.

\subsection{Spin-0 Sector}

The spin-0 sector consists of the sfermion and Higgs fields. For the sfermion fields let us begin with the up-type squark, as the masses for the other sfermions will be similar. The mass squared matrix for the up-type squark fields of the first family is given by

\begin{eqnarray}
\begin{pmatrix}
\t u_1 & \t {\bar{u}}_1^*
\end{pmatrix} \begin{pmatrix}
y_u^2 a^2/2 &  m b y_u/\sqrt{2} \\ m b y_u/\sqrt{2} & y_u^2 a^2/2
\end{pmatrix} \begin{pmatrix}
\t u_1^* \\ \t {\bar{u}}_1
\end{pmatrix},
\end{eqnarray}
where $\t u$ is the up-type squark in the $SU(2)$ doublet $\t Q$, and $a$ and $b$ are defined to be

\begin{eqnarray}
a &\equiv & \cos \theta_- v_1 - \sin \theta_- v_2, \nonumber \\
b &\equiv & \sin \theta_- v_1 + \cos \theta_- v_2.
\end{eqnarray}
Its eigenvalues are

\begin{eqnarray}
m_{\t u_1}^2 = \frac{y_u^2a^2}{2} \pm \frac{mby_u}{\sqrt{2}}.
\end{eqnarray}
For the second family, the mass squared matrix is

\begin{eqnarray}
\begin{pmatrix}
\t u_2 & \t {\bar{u}}_2^*
\end{pmatrix} \begin{pmatrix}
y_u^2 b^2/2 & (ma+bv) y_u/\sqrt{2} \\  (ma+bv) y_u/\sqrt{2} & y_u^2 b^2/2
\end{pmatrix} \begin{pmatrix}
\t u_2^* \\ \t {\bar{u}}_2
\end{pmatrix},
\end{eqnarray}
which has the eigenvalues

\begin{eqnarray}
m_{\t u_2}^2 = \frac{y_u^2b^2}{2} \pm \frac{(ma+bv) y_u}{\sqrt{2}}.
\end{eqnarray}
By replacing $y_u$ with $y_d$ and $y_e$ in the above formulae, one can obtain the masses for $\t d$ and $\t e$ respectively, while the sneutrino is massless at tree-level due to the absence of the right-handed sneutrino in this model. We note that because of the negative signs in the above mass formulae, the sfermion fields may have a negative mass squared in some regions of the parameter space.   

For the Higgs fields, we first note that there is no mass mixing terms between the $H_+$ and $H_-$ fields. Secondly, the $H_-$ fields contain the massless Nambu-Goldstone modes.

Let us start with the $H_+$ fields. Because $U(1)_\gamma$ is unbroken, the charged components of $H_+$ cannot mix with the neutral ones. Replacing $H_-$ with its $vev$, one finds the mass squared matrix for the charged components of $H_+$ to be

\begin{eqnarray}
\begin{pmatrix}
H_{+1}^+ & H_{+2}^+
\end{pmatrix} \begin{pmatrix}
m_{+1}^2+g^2\bar{a}^2/8-\eta\cos^2\theta_+ a^2/2 & g^2\bar{a}\bar{b}/8+\eta\sin(2\theta_+) a^2/4 \\ g^2\bar{a}\bar{b}/8+\eta\sin(2\theta_+) a^2/4 &
m_{+2}^2+g^2\bar{b}^2/8-\eta\sin^2\theta_+ a^2/2
\end{pmatrix} \begin{pmatrix}
H_{+1}^{+*} \\ H_{+2}^{+*}
\end{pmatrix}, \nonumber
\end{eqnarray} 
where $\bar{a}$ and $\bar{b}$ are defined as

\begin{eqnarray}
\bar{a} &\equiv & \cos \bar{\theta} v_1 + \sin \bar{\theta} v_2, \nonumber \\
\bar{b} &\equiv & -\sin \bar{\theta} v_1 + \cos \bar{\theta} v_2.
\end{eqnarray}

As one can imagine, the exact expressions for the eigenvalues of the above matrix are quite messy and not very illuminating. Instead, we choose to look at their expansions, which are given below in terms of the small parameters $x \equiv m/v$ and $y \equiv F/v^2$ by

\begin{itemize} 

\item $0<x^2 \ll y<1$:
\begin{eqnarray}
m_{H_{+1}^+}^2 &\approx & v^2 \left[ \frac{g^2}{4\eta} x^2 y -\left (2-\frac{g^2}{4\eta} \right) x^2y^2 +\m O(x^4) \right], \nonumber \\
m_{H_{+2}^+}^2 &\approx & v^2 \left[ 1+y +\m O( x^2) \right],
\end{eqnarray}
\item $0<x^2 \lesssim y<1$:
\begin{eqnarray}
m_{H_{+1}^+}^2 &\approx & v^2 \left[ 2 x^4+\frac{g^2}{4\eta}\left (1-\frac{x^2}{y} \right )x^4 +\m O\left ( \left (1-\frac{x^2}{y} \right )^2 x^4, x^6 \right) \right], \nonumber \\
m_{H_{+2}^+}^2 &\approx & v^2 \left[ 1+3 x^2 +\m O\left ( x^4, \left (1-\frac{x^2}{y} \right ) x^2 \right) \right],
\end{eqnarray}
where we have used $1-x^2/y$ as one of our expansion parameters.
\end{itemize}

The mass squared matrix for the neutral components can also be found and written as\footnote{The presence of an extra factor $1/2$ is because we have separated out all the real degrees of freedom of $H_+^0$.}

\begin{eqnarray}
\frac{1}{2}
\begin{pmatrix}
 H_+^0 & H_+^{0*}
\end{pmatrix} \begin{pmatrix}
\B A & \B B \\ \B B^t & \B A
\end{pmatrix} \begin{pmatrix}
H_+^{0} \\ H_+^{0*}
\end{pmatrix},
\end{eqnarray}
where $\B A$ and $\B B$ are matrices in family space given by

\begin{eqnarray}
\B A &=& 2\begin{pmatrix}
(g^2+g^{\prime 2})\bar{a}^2/32-\eta a^2\cos^2\theta_+/4 & (g^2+g^{\prime 2})\bar{a}\bar{b}/32 + \eta a^2 \sin(2\theta_+)/8 \\ (g^2+g^{\prime 2})\bar{a}\bar{b}/32 + \eta a^2 \sin(2\theta_+)/8 & (g^2+g^{\prime 2})\bar{b}^2/32-\eta a^2\sin^2\theta_+/4
\end{pmatrix}, \nonumber \\
\B B &=& \begin{pmatrix}
m_{+1}^2+(g^2+g^{\prime 2})\bar{a}^2/16-\eta a^2\cos^2\theta_+ & (g^2+g^{\prime 2})\bar{a}\bar{b}/16 + \eta a^2 \sin(2\theta_+)/2 \\ (g^2+g^{\prime 2})\bar{a}\bar{b}/16 + \eta a^2 \sin(2\theta_+)/2 & m_{+2}^2+(g^2+g^{\prime 2})\bar{b}^2/16-\eta a^2\sin^2\theta_+
\end{pmatrix}. \nonumber
\end{eqnarray}
Again, to get a feel for the eigenvalues of this complicated matrix, we look at their expansions:

\begin{itemize}

\item $0<x^2 \ll y<1$:
\begin{eqnarray}
m_{H_{+1}^0}^2 &\approx & v^2 \left[ \frac{g^2+g^{\prime 2}-8\eta}{4\eta}x^2y +\m O(x^4,x^2y^2) \right], \nonumber\\
m_{H_{+2}^0}^2 &\approx & v^2 \left[ x^4-x^2y^2+\m O ( x^6) \right], \nonumber \\
m_{H_{+3,4}^0}^2 &\approx & v^2 \left[ 1+y +\m O(x^2) \right],
\end{eqnarray}

\item $0<x^2 \lesssim y<1$:
\begin{eqnarray}
m_{H_{+1}^0}^2 &\approx & v^2 \left[ 2 x^4+\frac{g^2+g^{\prime 2}-8\eta}{4\eta}\left (1-\frac{x^2}{y} \right )x^4 +\m O\left ( \left (1-\frac{x^2}{y} \right )^2 x^4, x^6 \right) \right], \nonumber \\
m_{H_{+2}^0}^2 &\approx & v^2 \left[ 2 x^4+ +\m O\left (x^6,  \left (1-\frac{x^2}{y} \right ) x^6 \right) \right], \nonumber \\
m_{H_{+3,4}^0}^2 &\approx & v^2 \left[ 1+3 x^2 +\m O\left ( x^4, \left (1-\frac{x^2}{y} \right ) x^2 \right) \right],
\end{eqnarray}
\end{itemize}
where the last two eigenvalues for both cases are only equal up to the above orders.

For the $H_-$ fields, we begin by parametrizing our fields as 

\begin{eqnarray}
H_{- 1} &=& \begin{pmatrix}
\left(v_1+h_1 (x)\right) e^{i A \cos \beta /v_1} \\ H_{-1}^-
\end{pmatrix}, \\
H_{- 2} &=& \begin{pmatrix}
\left( v_2+h_2 (x) \right) e^{i A \sin \beta /v_2} \\ H_{-2}^-
\end{pmatrix},
\end{eqnarray}
in terms of the real fields $h_i$ and $A$, and complex fields $H_{-i}^-$. We find that the CP odd $A$ field has a mass given by

\beq
m_A^2 = \frac{\sin 2 \theta_- ( v_1^2 {\cos}^2 \theta_- + v_2^2 {\sin}^2 \theta_- )(1+\zeta^2)}{2\zeta},
\eeq
where $\zeta \equiv v_2 / v_1 $. 

The $H_-^-$ fields will mix, giving the following mass matrix

\beq 
\begin{pmatrix}
H_{-1}^- & H_{-2}^-
\end{pmatrix}
\begin{pmatrix}
m_{-1}^2 + \Pi^-_{11} & \Pi^-_{12} \\ \Pi^-_{12} & m_{-2}^2 + \Pi^-_{22}
\end{pmatrix}
\begin{pmatrix}
H_{-1}^{-*} \\ H_{-2}^{-*}
\end{pmatrix},
\eeq
where 
\bea
\Pi^-_{11} &=& \frac{\eta}{4} \left( 2 v_1^2 \cos^4 \theta_- + \frac{v_2^2}{2} \sin^2 2 \theta_- - 2 v_1 v_2 \sin 2 \theta_- \cos^2 \theta_- \right), \nonumber \\
\Pi^-_{12} &=& \frac{\eta}{4} \left(v_1 v_2 \sin^2 2 \theta_- - (v_1^2 \cos^2 \theta_- +v_2^2 \sin^2 \theta_-)\sin 2 \theta_- \right), \nonumber \\
\Pi^-_{22} &=&  \frac{\eta}{4} \left( 2 v_1^2 \sin^4 \theta_- + \frac{v_1^2}{2} \sin^2 2 \theta_- - 2 v_1 v_2 \sin 2 \theta_- \sin^2 \theta_- \right) \nonumber.
\eea
With this exact expression, one may verify that the above matrix has zero determinant, giving the Nambu-Goldstone bosons that are eaten by the gauge fields.  This leaves us with one massive negatively charged scalar. To get a feel for its mass, we expand as before, finding

\begin{itemize}
\item $0<x^2 \ll y<1$:
\beq
m_{H_-^-}^2 \approx v^2 \left[ 1 + 2 x^2 +(x^2-1) y + \m O \left(x^4 , x^2 y^2 \right) \right],
\eeq
\item $0<x^2 \lesssim y<1$:
\beq
m_{H_-^-}^2 \approx v^2 \left[ 1 + x^2 - x^2 \left(1-\frac{x^2}{y}\right) + \m O \left(x^8 , x^4 (1-\frac{x^2}{y})\right) \right].
\eeq 
\end{itemize}

For the neutral components $h_1$ and $h_2$ we find a similarly complicated mixing matrix,

\beq
\frac{1}{2}
\begin{pmatrix}
h_1 & h_2 
\end{pmatrix}
\begin{pmatrix}
2 m^2_{-1} +\Pi^0_{11} & \Pi^0_{12} \\ \Pi^0_{12} & 2 m^2_{-2} + \Pi^0_{22}
\end{pmatrix}
\begin{pmatrix}
h_1 \\ h_2 
\end{pmatrix},
\eeq 
where now 
\bea
\Pi^0_{11} &=& \frac{\eta}{2} \left( 6 v_1^2 \cos^4 \theta_- + \frac{3}{2} v_2^2 \sin^2 2 \theta_- - 6 v_1 v_2 \cos^2 \theta_- \sin 2 \theta_- \right), \nonumber \\
\Pi^0_{12} &=& \frac{\eta}{2} \left( 3 v_1 v_2 \sin^2 2 \theta_- - 3 \sin 2 \theta_- (v_1^2 \cos^2 \theta_- + v_2^2 \sin^2 \theta_-) \right), \nonumber \\
\Pi^0_{22} &=& \frac{\eta}{2} \left( 6 v_2^2 \sin^4 \theta_- + \frac{3}{2} v_1^2 \sin^2 2 \theta_- - 6 v_1 v_2 \sin^2 \theta_- \sin 2 \theta_- \right) \nonumber .
\eea
The eigenvalues are again complicated and not illuminating.  Expanding, we find a ``light'' and ``heavy'' Higgs field $h$ and $H$ 

\begin{itemize}
\item $0<x^2 \ll y<1$:
\bea
m^2_h &\approx & 2 v^2 \left[ 2 x^2 y -2 x^4 + \m O \left( x^6, x^4 y \right) \right], \\
m^2_H &\approx & 2 v^2 \left[ 1 + 2 x^2 -y + x^2 y  + \m O \left( x^6, x^4 y \right) \right],
\eea
\item $0<x^2 \lesssim y<1$:
\bea
m^2_h &\approx & 2 v^2 \left[ 2 x^4 \left(1-\frac{x^2}{y}\right) + \m O \left( x^8, x^6 \left(1-\frac{x^2}{y}\right) \right) \right], \\
m^2_H &\approx & 2 v^2 \left[ 1 + x^2  -x^2 \left(1-\frac{x^2}{y}\right) + \m O \left( x^8, x^4 \left(1-\frac{x^2}{y}\right) \right) \right].
\eea
\end{itemize}

\subsection{Spin-$\frac{1}{2}$ Sector}

The fermions are the gauginos, Higgsinos and the quark and lepton fields. The quarks will obtain masses from their Yukawa couplings as usual, while the charginos and neutralinos will be mixtures of the gauginos and Higgsinos. 

Replacing $H_{-1,2}$ by their vacuum values, we obtain the  quark and lepton masses,

\begin{eqnarray}
m_{u1} \simeq \frac{y_u}{\sqrt{2}}(1+x\zeta)v_1, \qquad m_{u2} \simeq \frac{y_u}{\sqrt{2}}(x+\zeta)v_1, \nonumber \\
m_{d1} \simeq \frac{y_d}{\sqrt{2}} (1-x\zeta)v_1, \qquad m_{d2} \simeq \frac{y_d}{\sqrt{2}} (x+\zeta)v_1, \nonumber \\
m_{e1} \simeq \frac{y_e}{\sqrt{2}} (1-x\zeta)v_1, \qquad m_{e2} \simeq \frac{y_e}{\sqrt{2}} (x+\zeta)v_1, \nonumber
\end{eqnarray}
and  the fermions of the first family have larger masses than those of the second family.

For the charginos, their mass matrix is of the Dirac type and given by

\begin{eqnarray}
\begin{pmatrix}
\t W^- & \t H_{d1}^- & \t H_{d2}^-
\end{pmatrix} \begin{pmatrix}
0 & ga/\sqrt{2} & gb/\sqrt{2} \\
-ga/\sqrt{2} & 0 & m \\
-gb/\sqrt{2} & m & v
\end{pmatrix} \begin{pmatrix}
\t W^+ \\ \t H_{u1}^+ \\ \t H_{u2}^+
\end{pmatrix},
\end{eqnarray}
which results in the following approximate masses,

\begin{itemize}

\item $0<x^2 \ll y<1$:
\begin{eqnarray}
m_{\t C_1} &\approx & v \left[ 1+x^2+\m O(x^4) \right ] \nonumber, \\
m_{\t C_{2,3}} &\approx & v \left [ \frac{g^2}{\eta} x\sqrt{y} +\m O (x^2,xy^{3/2}) \right ], 
\end{eqnarray} 

\item $0<x^2 \lesssim y<1$:
\begin{eqnarray}
m_{\t C_1} &\approx & v \left[ 1+x^2+\m O(x^4) \right ] \nonumber, \\
m_{\t C_2} &\approx & v \left [ \frac{g^2}{\eta} \left( 1-\frac{x^2}{y} \right)x^2 +\m O \left( \left( 1-\frac{x^2}{y} \right)^2 x^2 \right ) \right ], \nonumber \\
m_{\t C_3} &\approx & v \left [ x^2 +\frac{g^2}{\eta} \left( 1-\frac{x^2}{y} \right) x^2 +\m O \left( \left( 1-\frac{x^2}{y} \right)^2 x^2, x^4 \right ) \right ].
\end{eqnarray} 
\end{itemize}

On the other hand, the neutralinos have the following Majorana mass matrix:

\begin{eqnarray}
\frac{1}{2}\begin{pmatrix}
\t B & \t W^3 & \t H_{u1}^0 & \t H_{u2}^0 & \t H_{d1}^0 & \t H_{d2}^0
\end{pmatrix} \begin{pmatrix}
0 & 0 & g^\prime a/2 & g^\prime b/2 & g^\prime a/2 & g^\prime b/2 \\
0 & 0 & ga/2 & gb/2 & -ga/2 & -gb/2 \\
g^\prime a/2 & ga/2 & 0 & 0 & 0 & -m \\
g^\prime b/2 & gb/2 & 0 & 0 & -m & -v \\
g^\prime a/2 & -ga/2 & 0 & -m & 0 & 0 \\
g^\prime b/2 & -gb/2 & -m & -v & 0 & 0
\end{pmatrix} \begin{pmatrix}
\t B \\ \t W^3 \\ \t H_{u1}^0 \\ \t H_{u2}^0 \\ \t H_{d1}^0 \\ \t H_{d2}^0 
\end{pmatrix},\nonumber
\end{eqnarray}
and three of its mass eigenvalues are

\begin{itemize}

\item $0<x^2 \ll y<1$:
\begin{eqnarray}
m_{\t N_1} &\approx & v[1+x^2+\m O(x^4)] , \nonumber \\
m_{\t N_{2,3}} &\approx & v \left [ \frac{g}{\sqrt{\eta}} x\sqrt{y} +\m O (x^2) \right ], 
\end{eqnarray}

\item $0<x^2 \lesssim y<1$:
\begin{eqnarray}
m_{\t N_1} &\approx & v[1+x^2+\m O(x^4)] , \nonumber \\
m_{\t N_2} &\approx & v \left [ x^2 +\frac{g^2}{\eta} \left( 1-\frac{x^2}{y} \right) x^2 +\m O \left( \left( 1-\frac{x^2}{y} \right)^2 x^2, x^4 \right ) \right ], \nonumber \\
m_{\t N_3} &\approx & v \left [ \frac{g^2}{\eta} \left( 1-\frac{x^2}{y} \right)x^2 +\m O \left( \left( 1-\frac{x^2}{y} \right)^2 x^2, x^4 \right ) \right ],
\end{eqnarray}
\end{itemize}

while the other three can be obtained by replacing $g$ with $g^\prime$ in the above formulae.

The gluinos in this case will remain massless, as we have not included a color triplet Higgs and its family partner.  Although this is a problem phenomenologically, it would hopefully be addressed in a more complete model.

\subsection{Spin-1 Sector}

The mass pattern for the gauge fields is the same as in the Standard Model.  Three degrees of freedom from $H_{-}$ fields are eaten to give masses to the $W^\pm$ and $Z$ bosons, while the photon remains massless.


\section{Sfermion Spontaneous Symmetry Breaking}

In the aligned vacuum configuration when $m^2 > F$, negative sfermion masses squared are generated for most values of our parameters, signalling the spontaneous breaking of symmetries.  In order to get a feel for how such breaking occurs, we will for the time being neglect the squark contributions to the superpotential and focus on the sleptons. The slepton part of the scalar potential is  

\begin{eqnarray} \label{eqn:sleppot}
V_{\t l} &=& m_{\t L_1}^2|\t \nu_1|^2+m_{\t L_1}^2|\t e_1|^2+m_{\t e_1}^2|\t {\bar{e}}_1|^2+(1\rightarrow 2) \nonumber \\
  & & +|y_e|^2|\t \nu_{1}\t {\bar{e}}_1|^2+|y_e|^2|\t e_{1}\t {\bar{e}}_1|^2+(1\rightarrow 2) \nonumber \\
  & & +\frac{g^{\prime 2}}{8}[-|\t \nu_{1}|^2-|\t e_{1}|^2+2|\t {\bar{e}}_1|^2+(1\rightarrow 2)]^2 \nonumber \\
  & & +\frac{g^2}{8}[(\t e_{1}^*\t \nu_{1}+\t \nu_{1}^*\t e_{1})+(1\rightarrow 2)]^2 \nonumber \\
  & & -\frac{g^2}{8}[(\t e_{1}^*\t \nu_{1}-\t \nu_{1}^*\t e_{1})+(1\rightarrow 2)]^2 \nonumber \\
  & & +\frac{g^2}{8}[(\t \nu_{1}^*\t \nu_{1}-\t e_{1}^*\t e_{1})+(1\rightarrow 2)]^2,
\end{eqnarray}
where $\t \nu$ and $\t e$ form the left handed $SU(2)$ doublet $\t L$.
 
We consider the limit $0 < y \ll x^2 \lesssim 1$.  As can be seen from Eq.(\ref{eqn:example}) or Fig.\ref{fg:sfermion}, this case is perhaps the simplest for exploring our question, as to leading order the sfermions of both families have the same masses squared but with opposite signs, the first families being negative. Minimization of the above potential with respect to the second family, $\langle \partial V_{\t l}/\partial \t l_2 \rangle =0$, implies that the second family receives no vacuum value: $\langle \t \nu_{L2} \rangle = \langle \t e_{L2} \rangle =\langle \t {\bar{e}}_2 \rangle=0$. All symmetry breaking will be done by the first family. 

Our initial symmetries are $\m S_3 \times SU(2)\times U(1)_Y \times U(1)_{l}$. Due to the $SU(2)$ invariance, one can choose

\begin{eqnarray}
\langle \t L_1 \rangle = \langle \begin{pmatrix}
\t \nu_{1} \\ \t e_{1}
\end{pmatrix} \rangle = \begin{pmatrix}
\frac{v_1}{\sqrt{2}} \\ 0
\end{pmatrix}, \quad\quad \langle \t {\bar{e}}_1 \rangle= \frac{v_2}{\sqrt{2}},
\end{eqnarray}
so that the vacuum value of $\t L_1$ leaves an unbroken $U(1)_{I_3 + Y/2} \equiv U(1)_\gamma$. To examine which symmetries are spontaneously broken and which are left unbroken, we analyse separately the case when only $v_1$ is non-zero, only $v_2$ is non-zero, or neither is non-zero. Since the first pattern of symmetry breaking leaves unbroken a gauged $U(1)_{\gamma}$ associated with the electromagnetic charge, we can compare our results to the usual spectrum achieved in the MSSM after SSB, and so we shall study it in greater detail.

\begin{enumerate}
\item
If only $v_1$ is non-zero, the equation $\langle \partial V_{\t l}/\partial \t L_1 \rangle =0$ gives the solution

\begin{equation}
\frac{v_1}{\sqrt{2}} = \frac{2 |m_{\t L_1}|}{\sqrt{g^2+g'^2}}.
\end{equation}

To obtain the physical spectrum, we expand our fields away from this minimum in the unitary gauge,

\begin{equation} \label{eq:expansion1}
\t L_1 = e^{i (\xi^a(x) \tau^a )/v_1} \begin{pmatrix}
\frac{1}{\sqrt{2}} (v_1 +\t \nu_1(x)) \\ 0 
\end{pmatrix} \equiv \m U(x) \begin{pmatrix}
\frac{1}{\sqrt{2}} (v_1 +\t \nu_1(x)) \\ 0 \end{pmatrix} .
\end{equation}
Substituting this back into Eq.(\ref{eqn:sleppot}), we find that the $\xi$ fields disappear, to be eaten by the gauge bosons, and the part of the potential including only the real field $\t \nu_1 (x)$ is given by 

\begin{equation}
V_{\t \nu_1} = -\frac{2}{g^2+g'^2}(m_{\t L_1}^2)^2 + \frac{1}{2}\t \nu_1^2(x)(2 m_{\t L_1}^2) + \frac{1}{8} (g^2+g^{\prime 2})v_1 \t \nu_1^3(x) +\frac{1}{32} (g^2+g^{\prime 2}) \t \nu_1^4(x).
\end{equation} 
The potential now has a minimum value given by 
\begin{equation}
V_{min} = -\frac{2}{g^2+g'^2}(m_{\t L_1}^2)^2.
\end{equation}
We next look at the mass spectrum for the various other fields. For the scalar fields, in addition to the radiative corrections obtained in the previous section, new contributions due to the above vacuum appear:

\begin{itemize}
\item
A contribution of $(y_e^2 v_1^2)/2$ for $\t{\bar{e}}_1$ and $\m H_{d 1}^0$ from their Yukawa interactions with $\t L_1$.
\item
A contribution from the D terms of $-\frac{g^{\prime 2}}{8} Y_{\varphi} v_1^2$ to all scalar fields $\varphi$ with non-zero hypercharge except $\t \nu_1(x)$. 
\item 
A contribution from the D terms of $\frac{g^2}{4} (I_3)_{\varphi} v_1^2$ to any neutral component of an SU(2) doublet except $\t \nu_1(x)$. 
\end{itemize}
For the fermion fields, both the Yukawa interactions in the superpotential,

\begin{equation}
\int d^2 \theta y_e L_1 \bar{e}_1 \m H_{d 1} +\ldots \rightarrow y_e \t L_1 \bar{e}_1 \t{\m H}_{d 1} +\ldots
\end{equation}
and the supersymmetric gauge interactions

\begin{equation}
\sqrt{2} g \t L_1^* \frac{\tau^a}{2} L_1 \t W^a - \frac{\sqrt{2}}{2} g^\prime \t L_1^*  L_1 \t B +\ldots
\end{equation}
will lead to various mass mixing terms. With the original gaugino and higgsino mass terms considered, this mixing occurs in three distinct sectors. 

In the first, we find Dirac type masses between the fields $\bar{e}_1$, $\t{\m H}_{u 1}^+$, $\t{\m H}_{u 2}^+$, and $\t{\m H}_{d 1}^-$, $\t{\m H}_{d 2}^-$. The mass matrix, given by 

\begin{equation}
\begin{pmatrix}
\bar{e}_1 & \t{\m H}_{u 1}^+ & \t{\m H}_{u 2}^+ 
\end{pmatrix} \begin{pmatrix}
\frac{y_e v_1}{\sqrt{2}} & 0 \\
 0 & m \\
 m & v 
\end{pmatrix} \begin{pmatrix}
\t{\m H}_{d 1}^- \\  \t{\m H}_{d 2}^-
\end{pmatrix}, \nonumber
\end{equation}
can be diagonalized through a singular value decomposition, leading to three fermionic mass eigenstates:

\begin{itemize}
\item
One massless Weyl fermion,
\begin{equation}
\t C_0 \equiv \frac{1}{\sqrt{m^4 + ((y_e v_1)^2 / 2) (m^2 + v^2)}}\left(m^2 \bar{e}_1 + \frac{y_e v_1}{\sqrt{2}} v \t{\m H}_{u 1}^+ - \frac{y_e v_1}{\sqrt{2}} m \t{\m H}_{u 2}^+ \right). \nonumber
\end{equation}

\item Two massive Dirac fermions, which we call $\t C_{1,2}$, of mass 
\begin{equation}
\frac{1}{\sqrt{2}} \sqrt{\frac{(y_e v_1)^2}{2} + 2m^2 + v^2 \pm \sqrt{\left( \frac{(y_e v_1)^2}{2} -v^2 \right)^2 + 4 m^2 v^2} }. \nonumber
\end{equation}
\end{itemize}

The second sector consists of the mixing between the charged components of $\t W^a$, 

\begin{equation}
\t W^\pm = \frac{1}{\sqrt{2}}\left(\t W^1 \mp i \t W^2 \right), \nonumber
\end{equation}
and the left handed electron $e_1^-$. The mass term for this sector is also of the Dirac type, 

\begin{equation}
\t W^+ \left( M_2 \t W^- +\frac{g v_1}{\sqrt{2}} e_1^- \right), \nonumber
\end{equation}
and here we find:
\begin{itemize}
\item
A massless Weyl fermion given by

\begin{equation}
\t C^{\prime}_0 \equiv \frac{1}{\sqrt{((g v_1)^2)/2+ M_2^2}}\left( - g \frac{v_1}{\sqrt{2}} \t W^- + M_2 e_1^- \right) \nonumber.
\end{equation}
\item
One massive Dirac fermion, which we call $\t C^{\prime}_{1}$, of mass
\begin{equation}
\sqrt{M_2^2 + \left(g \frac{v_1}{\sqrt{2}} \right)^2} . \nonumber
\end{equation}
\end{itemize}

The third sector involves the neutral fermions $\t W^3$, $\t B$, and $\nu_1$.  They mix through a Majorana type mass matrix given by

\begin{equation}
\frac{1}{2}
\begin{pmatrix}
\t B & \t W^3 & \nu_1 
\end{pmatrix}
\begin{pmatrix}
M_1 & 0 & - g^\prime v_1 \\
0 & M_2 & g v_1 \\
-g^\prime v_1 & g v_1 & 0 \\
\end{pmatrix}
\begin{pmatrix}
\t B \\ \t W^3 \\ \nu_1 
\end{pmatrix}. \nonumber
\end{equation}
The determinant of the above matrix does not vanish, signalling that we have three massive neutralinos, $\t N_{1, 2, 3}$. The mass formulae for these particles are not very enlightening, but the useful relation 

\begin{equation}
\t N_1 + \t N_2 + \t N_3 = M_1 + M_2 
\end{equation}
tells us that their average mass is around that of the average of the gaugino soft masses. 

The gauge bosons acquire masses through the quadratic terms in the covariant derivative as usual,

\begin{equation}
|\m D_\mu \t L_1|^2 = \left| \frac{\m U(x)}{\sqrt{2}} \begin{pmatrix} \partial_\mu \t \nu_1(x) + \frac{i}{2}(g W^3_\mu - g^\prime B_\mu)(v_1 + \t \nu_1(x)) \\ \frac{i}{2} g(W^1_\mu + i W^2_\mu )(v_1  + \t \nu_1(x)) \end{pmatrix} \right|^2 . 
\end{equation}
This leaves us with the following spectrum:

\begin{itemize} 
\item A massive real scalar, $\t \nu_1(x)$, with mass $2 |m_{\t L_1}^2|$.

\item Three massive gauge bosons, $W^{\pm}_\mu$ and $Z_\mu$, exactly as in the SM, with masses $\frac{1}{4} v_1^2 g^2$ and $\frac{1}{4} v_1^2 (g^2 + g^{\prime 2})$, respectively.

\item A massless gauge boson, $A_\mu$, corresponding to the unbroken $U(1)_{\gamma}$.

\end{itemize}

In this vacuum $SU(2)$ is spontaneously broken, as well as one of the $U(1)$'s, but the two linear combinations $U(1)_{2I_3 -l} \times U(1)_{\gamma}$ remain unbroken. The second is gauged, the first a left over global symmetry.  By looking at the values of the various fields under this global symmetry (see Table \ref{tb:charges}), we see that this global symmetry dictates the allowed mixing between the fields, and eliminates the hope of forming mass terms (of both Dirac and Majorana types) for these two massless Weyl fermions.

In summary, although the symmetry breaking pattern and gauge boson sector is similar to that of the SM, the mass spectrum generated in the lepton sector is quite different.  Instead of a massless neutral particle, which we could hope to identify with the left handed neutrino say, we find that all neutral fermions are massive with roughly the same order. The massless particles are found instead in the charged sector. One consequence of this is that the charged current associated with a $W$ boson contains a \emph{massless} charged fermion and a \emph{massive} neutral fermion, the opposite of the situation in the SM.   

\item
When only $v_2$ is non-zero, we find instead the solutions (using the relation $m_{\t{\bar{e}}_1}^2 = 2 m_{\t L_1}^2$)

\begin{eqnarray}
\frac{v_2}{\sqrt{2}} &=& \frac{|m_{\t{\bar{e}}_1}|}{g'} \nonumber \\
V_{min} &=& - \frac{2}{g^{\prime 2}}(m_{\t L_1}^2)^2.
\end{eqnarray}
Expanding around this minimum, 

\begin{equation}
\t{\bar{e}}_1 = e^{i \zeta(x)/v_2} \left(
\frac{1}{\sqrt{2}} (v_2 +\t{\bar{e}}_1(x))\right) \equiv \m U(x) \left(
\frac{1}{\sqrt{2}} (v_2 +\t{\bar{e}}_1(x))\right) ,
\end{equation}
the scalar potential for the real field $\t{\bar{e}}_1(x)$ is given by

\begin{equation}
V_{\t{\bar{e}}_1} = -\frac{2}{g'^2}(m_{\t L_1}^2)^2 + \frac{1}{2}\t{\bar{e}}_1^2(x)(4 m_{\t L_1}^2) + \frac{1}{2} g^{\prime 2} v_2 \t{\bar{e}}_1^3(x) +\frac{1}{8} g^{\prime 2} \t{\bar{e}}_1^4(x),
\end{equation}
and expanding the covariant derivative

\begin{equation}
|\m D_\mu \t{\bar{e}}_1|^2 = \left|\frac{\m U(x)}{\sqrt{2}} \left( \partial_\mu \t{\bar{e}}_1(x) + i g^{\prime} B_\mu (v_2 +\t{\bar{e}}_1(x)) \right) \right|^2,
\end{equation}
our spectrum now consists of

\begin{itemize}

\item A massive real scalar field, $\t{\bar{e}}_1(x)$ with mass $2 m_{\t{\bar{e}}_1}^2 = 4 m_{\t L_1}^2$.

\item Three massless gauge bosons $W^a_\mu$, $a= 1,2,3$, corresponding to the unbroken gauged $SU(2)$. 

\item A massive gauge boson, $B_\mu$, with mass $m_{\t{\bar{e}}_1}^2$.

\end{itemize}

In addition to the above masses, the mass spectrum of other fields can be obtained as well.  However, since the symmetry breaking pattern makes this case difficult to compare with a realistic model, we will leave the analysis of the spectrum here.

The unbroken symmetries are $SU(2) \times U(1)_{Y+2l}$, the first one gauged, the second a global symmetry.  Since $g^2$ is always greater than zero, we find that the second case has a deeper vacuum and is therefore always preferred to the first. 

\item
If both $v_1$ and $v_2$ are non-zero, the situation is more complicated but straightforward. A solution to the minimization equations gives the following vacuum values for the fields,

\begin{eqnarray}
\frac{v_1^2}{2} &=&  \frac{2 m_{\t L_1}^2 (g^{\prime 2} - y_e^2)}{y_e^4 - y_e^2 g^{\prime 2} - \frac{1}{4} g^2 g^{\prime 2}} \nonumber \\
\frac{v_2^2}{2} &=&  \frac{ m_{\t L_1}^2 (g^{\prime 2} + \frac{1}{2}g^2 + y_e^2)}{y_e^4 - y_e^2 g^{\prime 2} - \frac{1}{4} g^2 g^{\prime 2}} \nonumber \\
V_{min} &=& \frac{2 (m_{\t L_1}^2)^2 (g^{\prime 2} + \frac{1}{4}g^2 - y_e^2)}{y_e^4 - y_e^2 g^{\prime 2} - \frac{1}{4} g^2 g^{\prime 2}} .
\end{eqnarray}

Whether or not this minimum is deeper than when only one of $v_1$ or $v_2$ is non-zero now depends on the values of $|y_e|^2$, $g^2$ and $g^{\prime 2}$. When $|y_e|^2 \ll (g^2, g^{\prime 2})$, this is indeed the case, and the opposite is true if $|y_e|^2 \gg (g^2, g^{\prime 2})$. This is quite interesting, as we may have different phases of our model depending on the relative strengths of these parameters. 

To find the spectrum in the case $|y_e|^2 \ll (g^2, g^{\prime 2})$, we expand $\t L_1$ and $\t{\bar{e}}_1$ exactly as before, and the analysis is unchanged except for a complication in the masses of the neutral gauge bosons.  From the above covariant derivatives, we find the mass matrix is given by

\begin{equation}
\frac{1}{2} \begin{pmatrix}
B_\mu & W^3_\mu 
\end{pmatrix}
\begin{pmatrix}
\frac{g^{\prime 2}}{4}(v_1^2+ 4 v_2^2) & -\frac{g g^\prime}{4} v_1^2 \\
-\frac{g g^\prime}{4} v_1^2 & \frac{g^2}{4}v_1^2 
\end{pmatrix}
\begin{pmatrix}
B^\mu \\ W^{3 \mu}
\end{pmatrix},
\end{equation}
which has eigenvalues 

\begin{equation}
m_{Z^0_\pm}^2 = \frac{1}{2} \left( g^2 + g^{\prime 2} \right) v_1^2 + 2 \left( g^{\prime 2} v_2^2 \pm \sqrt{\left( \frac{1}{16} (g^2+g^{\prime 2})v_1^4 +g^{\prime 4} v_2^4 + 2 (g^{\prime 4} - g^2 g^{\prime 2}) v_1^2 v_2^2 \right)} \right).
\end{equation}
Our spectrum contains

\begin{itemize}

\item A massive real scalar, $\t \nu_1(x)$, with mass $2 m_{\t L_1}^2$.

\item A massive real scalar field, $\t{\bar{e}}_1(x)$ with mass $2 m_{\t{\bar{e}}_1}^2 = 4 m_{\t L_1}^2$.

\item Four massive gauge bosons, $W^\pm_\mu$ and $Z^0_{\pm \mu}$, with the masses given above.

\end{itemize}

In this case, the gauged $SU(2)$ and $U(1)$ are totally broken, leaving behind one global symmetry, $U(1)_{2I_3 - Y - 2 l}$. Again, slepton and lepton masses and mixings are generated, but we do not give the full spectrum here.  We summarize the various cases in Table \ref{tb:vacuum}, and the charges of our fields under the global symmetries in Table \ref{tb:charges}.
\end{enumerate}

\begin{table}[!h]
\begin{center}
\begin{tabular}{|c|c|c|c|c|}
\hline 
\hline
Case & $\langle \t \nu_{L1} \rangle$ & $\langle \t {\bar{e}}_1 \rangle$ & Residual symmetries  (gauge $\times$ global) & $V_{min}$ \\
\hline
1 & $v_1$ & $0$ & $U(1)_{\gamma} \times U(1)_{2I_3-l}$ & $-\frac{2}{g^2+g^{\prime 2}}(m_{\t L1}^2)^2$ \\
\hline
2 & $0$ & $v_2$ & $SU(2) \times U(1)_{Y+2l}$ & $-\frac{2}{g^{\prime 2}}(m_{\t L1}^2)^2$ \\
\hline
3 & $v_1$ & $v_2$ & $U(1)_{2I_3-Y-2l}$ ~(global) & $\frac{2 (m_{\t L_1}^2)^2 (g^{\prime 2} + \frac{1}{4}g^2 - y_e^2)}{y_e^4 - y_e^2 g^{\prime 2} - \frac{1}{4} g^2 g^{\prime 2}}$ \\
\hline
\hline
\end{tabular}
\caption{Different vacuum configurations of the sleptons. $V_{min}$ is the minimum of the potential.}
\label{tb:vacuum}
\end{center}
\end{table}

\begin{table} [!h]
\begin{center}
\begin{tabular}{|c|c|c|c|c|c|c|c|c|}
\hline
\hline
Case & Global Symmetry & $q_{\nu}$ & $q_{e}$ & $q_{\bar{e}}$ & $q_{\m H_u^+}$ & $q_{\m H_u^0}$ & $q_{\m H_d^-}$& $q_{\m H_d^0}$ \\
\hline
1 & $U(1)_{2I_3-l}$ & 0 & -2 & 1 & 1 & -1 & -1 & 1 \\
\hline
2 & $U(1)_{Y+2l}$ & 1 & 1 & 0 & 1 & 1 & -1 & -1  \\
\hline
3 & $U(1)_{2I_3-Y-2l}$  & 0 & -2 & 0 & 0 & -2 & 0 & 2 \\
\hline
\hline
\end{tabular}
\caption{Charges of the various lepton superfields under the unbroken global symmetries.}
\label{tb:charges}
\end{center}
\end{table}

For almost all other values of $x$ and $y$ satisfying $0<y<x^2<1$, the above analysis will be much the same as above. In some cases the roles of the families may be reversed, or the two families soft masses may not simply differ by a sign, but this will affect only the magnitudes and labelling of the $vev$'s.  One complication for values such as those shown in the first and last figures of Fig.\ref{fg:sfermion} is that both families have negative masses squared.  Even in this case, the pattern of symmetry breaking will be qualitatively the same as that given above, up to a monstrous exercise in the minimization of potentials. 
\newpage

\section{Soft Mass RG Figures}

We give the figures for the RG running of the soft masses for $v=10^8$ $GeV$, $m_{-1} = 10^2$ $GeV$, $m_{\t H} = 2 \times 10^6$ $GeV$,  $y_u = 0.96$, $y_d = y_e = 0.1$ , $g^\prime = 0.39$, $g = 0.65$, and $g_s = 1.02$. The peculiar behavior of the right handed charged sleptons in Fig.\ref{fg:Lrunning}, where their masses squared run negative at low energies, is due to the positive term proportional to $S \equiv \Tr [ Y m^2 ]$ in their RG equations. We believe this term will be less significant in a more realistic model. 

\begin{figure}[!h]
 \centering
\scalebox{0.4}{\includegraphics*{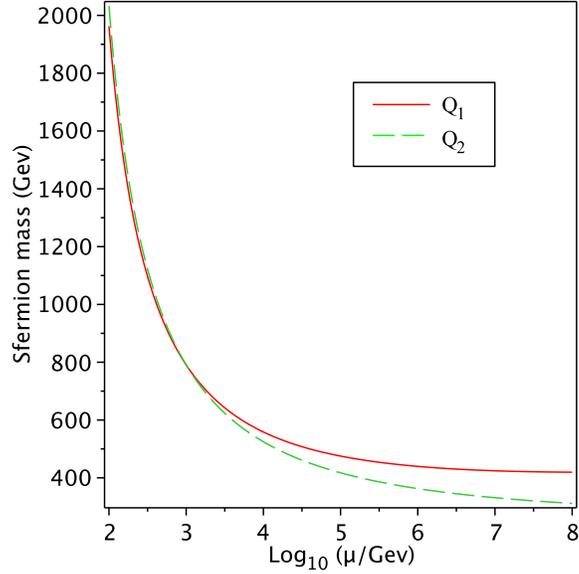}}
\caption{RGE running of $m_{\t Q_i}$.}
\label{fg:Qrunning}
\end{figure}

\begin{figure}[!h]
 \centering
\scalebox{0.4}{\includegraphics*{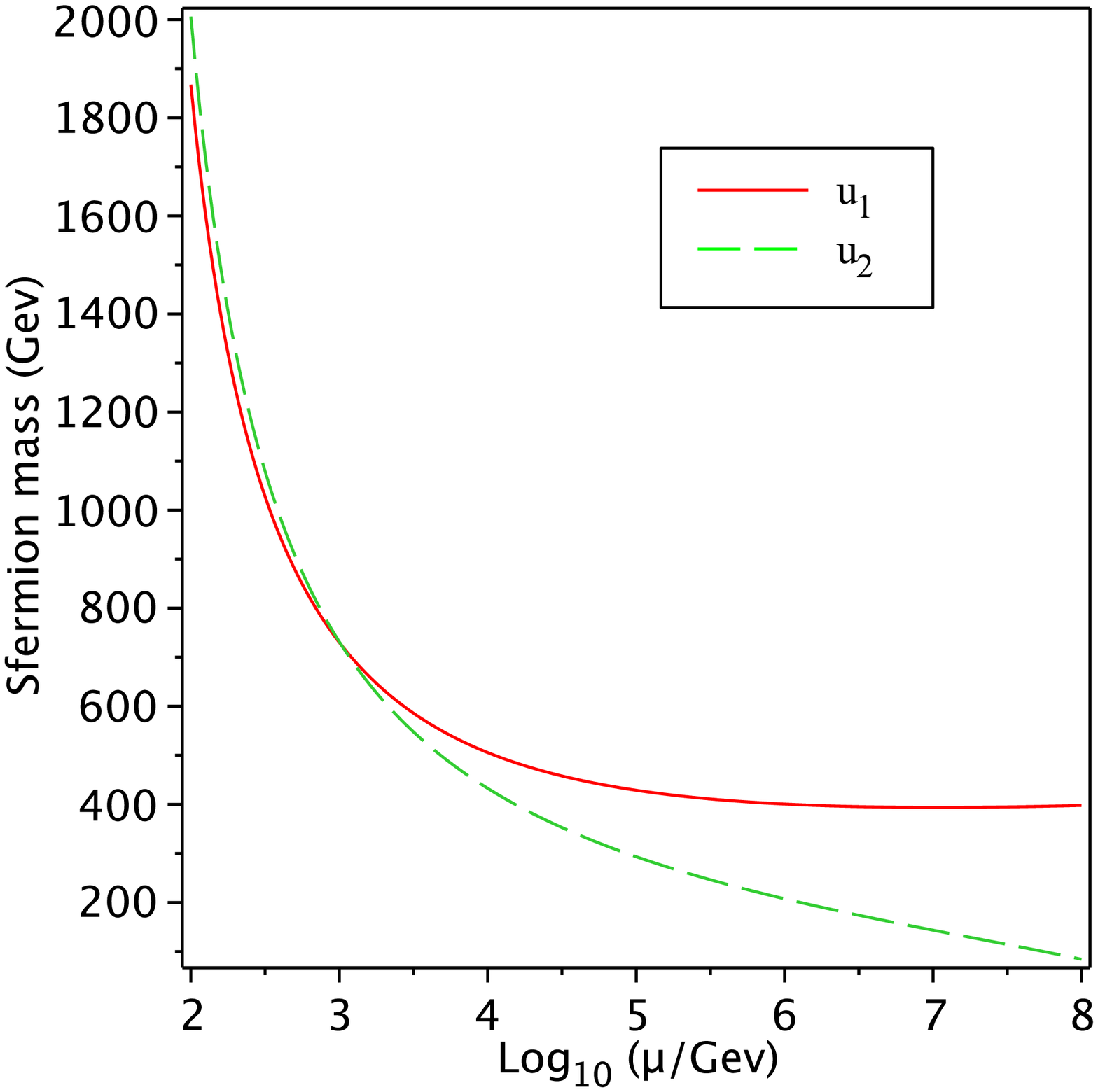}}
\caption{RGE running of $m_{\t{\bar{u}}_i}$.}
\end{figure}

\newpage
\begin{figure}[!h]
 \centering
\scalebox{0.45}{\includegraphics*{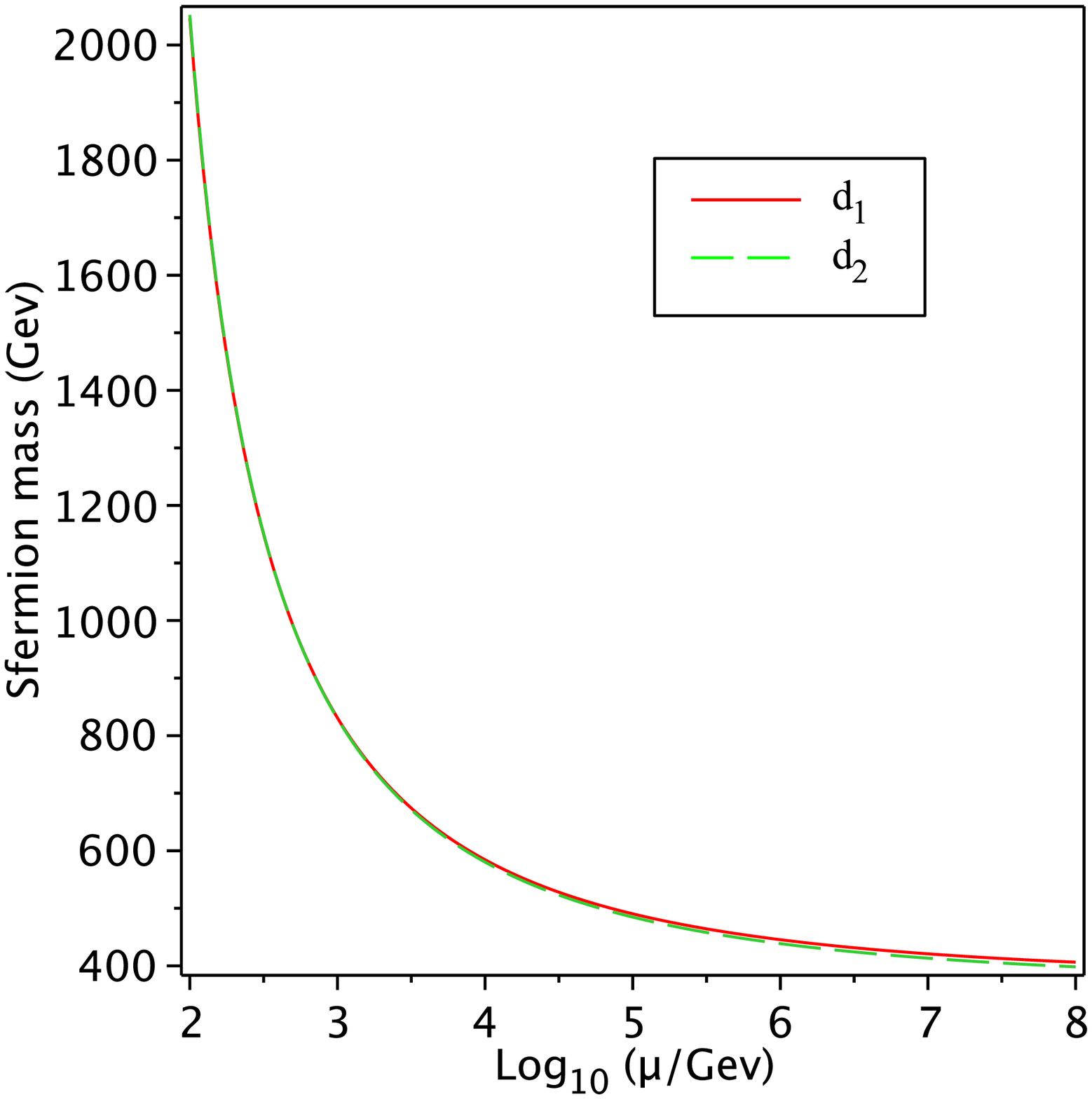}}
\caption{RGE running of $m_{\t{\bar{u}}_i}$.}
\label{fg:Drunning}
\end{figure}

\begin{figure}[!h]
 \centering
\scalebox{0.45}{\includegraphics*{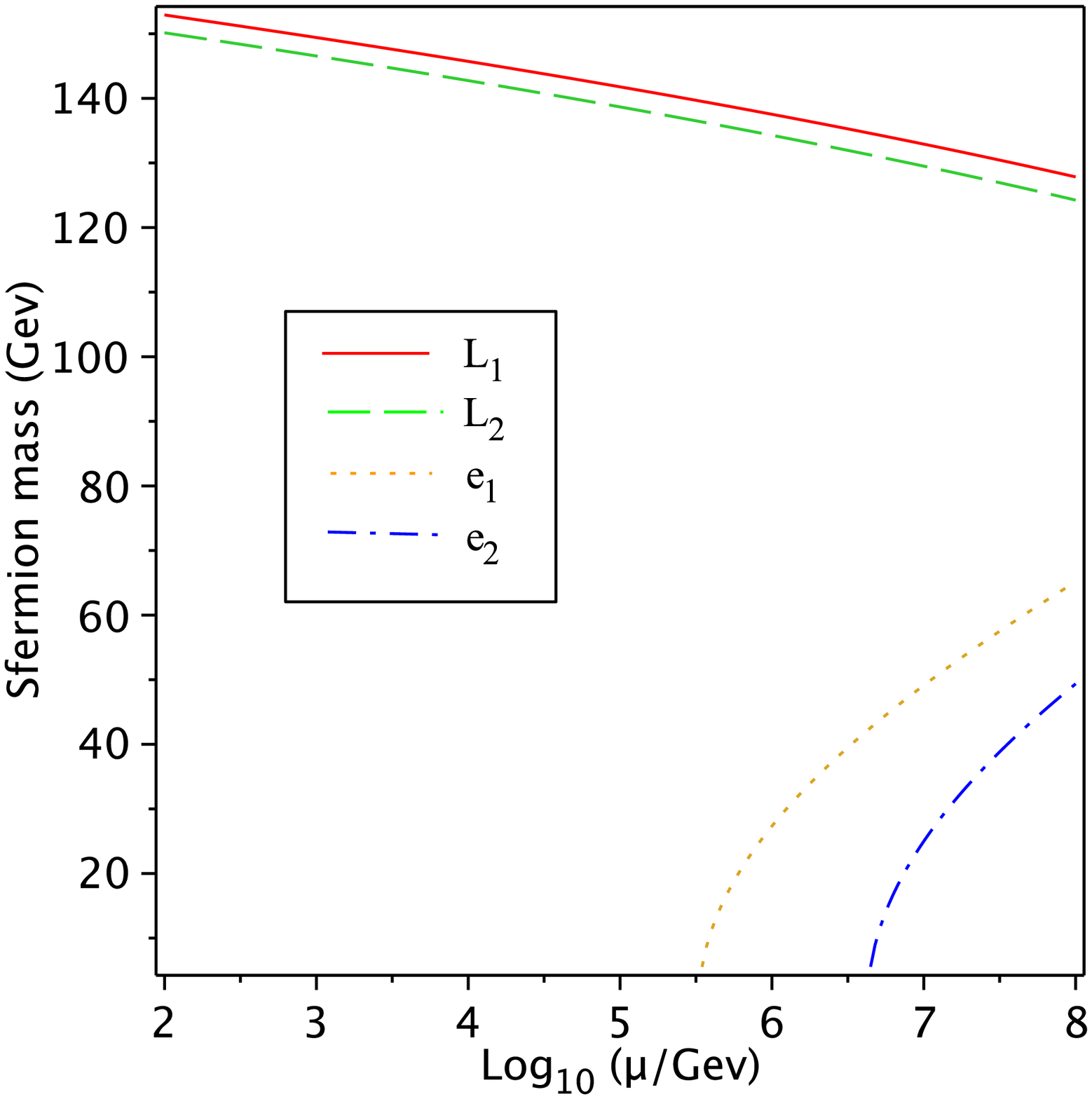}}
\caption{RGE running of $m_{\t L_i}$.}
\label{fg:Lrunning}
\end{figure}

\newpage
\begin{figure}[!h]
 \centering
\scalebox{0.45}{\includegraphics*{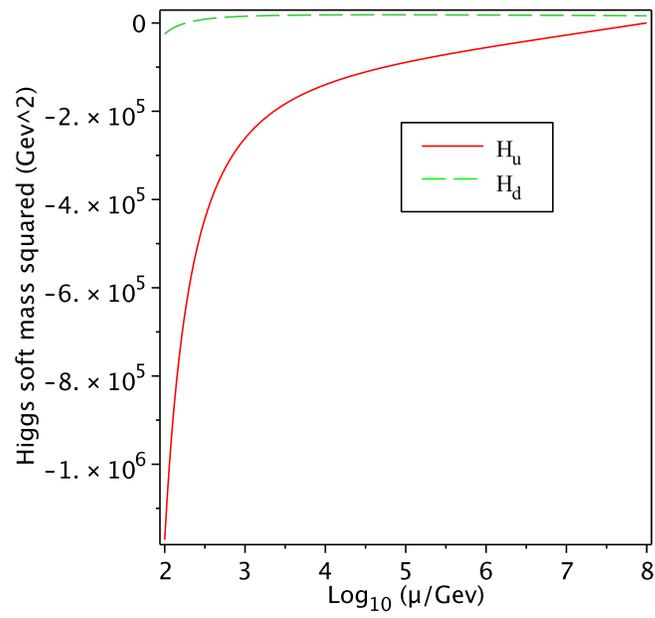}}
\caption{RGE running of $m_{H_u}^2$ and $m_{H_d}^2$.}
\label{fg:Hrunning}
\end{figure}

\begin{figure}[!h]
 \centering
\scalebox{0.45}{\includegraphics*{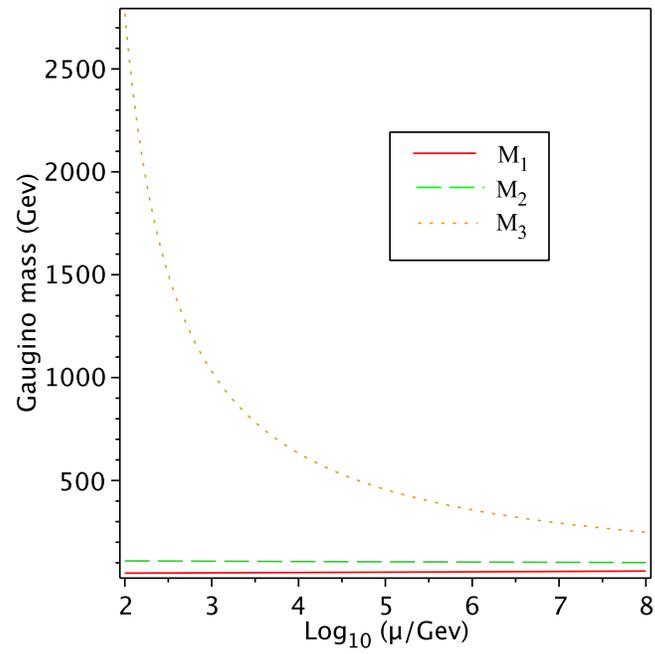}}
\caption{RGE running of $M_a$.}
\label{fg:Mrunning}
\end{figure}
\newpage


\end{document}